\documentclass{ptephy_v1}
\usepackage{hyperref}
\usepackage{amsmath,amsfonts,amssymb,mathtools,bm}
\usepackage{graphicx,subfig,ulem}
\usepackage{caption}
\newcommand{\Li}{\operatorname{Li}}

\begin{document}

\title{Convergence of Ginzburg--Landau Expansions: \\
Superconductivity in the Bardeen--Cooper--Schrieffer Theory and \\
Chiral Symmetry Breaking in the Nambu--Jona-Lasinio Model}

\author[1]{William Gyory}
\author[2]{Naoki Yamamoto}

\affil[1]{Graduate Center, City University of New York, 365 5th Avenue, New York, NY 10016, USA}

\affil[2]{Department of Physics, Keio University, Yokohama 223-8522, Japan}

\begin{abstract}%
We study the convergence of the Ginzburg--Landau (GL) expansion in the context of the Bardeen--Cooper--Schrieffer (BCS) theory for superconductivity and the Nambu--Jona-Lasinio (NJL) model for chiral symmetry breaking at finite temperature $T$ and chemical potential $\mu$. We present derivations of the all-order formulas for the coefficients of the GL expansions in both systems under the mean-field approximation. We show that the convergence radii for the BCS gap $\Delta$ and dynamical quark mass $M$ are given by $\Delta_{\rm conv} = \pi T$ and $M_{\rm conv} = \sqrt{\mu^2 + (\pi T)^2}$, respectively. 
We also discuss the implications of these results and the quantitative reliability of the GL expansion near the first-order chiral phase transition.
\end{abstract}

\subjectindex{A40, D30, I68}

\maketitle

\section{Introduction}
Power series expansions are ubiquitous in physics. Some examples can be found in the perturbative expansions appearing throughout quantum mechanics and quantum field theory (QFT). Another example is the paradigm of effective field theories (EFT), based on a systematic expansion at certain scales. Hydrodynamics, for instance, is an EFT based on a gradient expansion. Ginzburg--Landau (GL) theory, originally introduced in 1950 as a phenomenological model of superconductivity, is also an EFT based on the expansion of the free energy of a system in powers of an order parameter near a phase transition. 

It is well known that these power series do not always converge, and that even divergent series can generate useful results. Indeed, although the perturbative expansion in quantum electrodynamics (QED) should be a divergent asymptotic series with zero radius of convergence, as shown by Dyson \cite{Dyson:1952tj}, certain quantities calculated using perturbation theory, such as the anomalous magnetic dipole moment of the electron, are among the most precise predictions in physics. Several arguments have been put forward for the divergence of perturbative expansions in other types of QFTs, such as for scalar $\lambda\phi^3$ theory \cite{Hurst:1952zk}, and by 't Hooft for quantum chromodynamics (QCD) \cite{tHooft:1977xjm}. 
While convergent and divergent series can both be useful, there remain reasons---both practical and theoretical---for studying their convergence properties. On the practical side, it is important to know whether adding more terms to an expansion will produce better results, and if so, over what region of parameter space. On the theoretical side, at least in some situations, one can gain physical insight from an expansion's convergence or lack thereof. For recent developments on  asymptotic series in QFTs, see, e.g.\ the review in Ref.~\cite{Aniceto:2018bis}.
On the other hand, convergence properties of the systematic expansions for EFTs have yet to be understood generally. In this context, it was recently shown, using holographic duality methods, that the derivative expansions for specific physical quantities (shear and sound frequencies) in hydrodynamics have a finite radius of convergence in an $\mathcal N = 4$ supersymmetric Yang--Mills plasma~\cite{Grozdanov:2019kge}; see also Ref.~\cite{Cartwright:2021qpp} for its extension to rotating plasmas.

In this paper, we study the convergence of the GL expansion. As paradigmatic examples, we consider the Bardeen--Cooper--Schrieffer (BCS) theory for superconductivity and the Nambu--Jona-Lasinio (NJL) model for chiral symmetry breaking \cite{Nambu:1961tp} at finite temperature $T$ and chemical potential $\mu$. 
Despite the physical differences between the BCS superconductivity,  characterized by the BCS gap $\Delta$, and the chiral symmetry breaking in the NJL model, characterized by the dynamical quark mass $M$, their free energies in the mean-field approximation are closely related to a common mathematical expression, namely, 
\begin{align}
    \label{J_l}
    J_{\ell}(x, y) &=
    \int_0^\infty d t \,
    t^{\ell}
    \bigg[
    \sqrt{x^2 + t^2} + 
    \sum_{\zeta = \pm} \ln
    \left(
        1+e^{-\sqrt{x^2 + t^2} + \zeta y}
    \right)
    \bigg] \nonumber \\
    &= \int_0^\infty d t \,
    t^{\ell} \sum_{\zeta = \pm} \ln \bigg[ 2 \cosh
    \bigg(
        \frac
        {\sqrt{x^2 + t^2} + \zeta y} 2 
    \bigg) \bigg]\,,
\end{align}
where either $\ell=0$ or $\ell=2$. This expression must be regarded only schematically, because the first term in Eq.~(\ref{J_l}) is divergent and requires regularization. We will explain how $J_{\ell}$ appears in each context, and then we will show that the GL expansion essentially reduces to expanding Eq.~(\ref{J_l}) in powers of $x$, subject to some form of regularization. We will then derive the $n$th-order coefficients of the GL expansions for arbitrary $n$ in both systems.%
\footnote{Although the first few GL coefficients are well known in the BCS theory (see, e.g.\ Refs.~\cite{Gorkov1959, ADG}), the generic higher-order GL coefficients do not seem to be widely known, except for an unpublished note~\cite{Brauner}. To the best of our knowledge, the generic GL coefficients in the NJL model in $3+1$ dimensions and the radii of convergence of the GL expansions in both systems are not provided in the literature.}
We will show that the radius of convergence in each case is given by $\Delta_{\rm conv} = \pi T$ and $M_{\rm conv} = \sqrt{\mu^2 + (\pi T)^2}$, respectively, and we will clarify the physical origin of the difference between these two formulas. The finite convergence radii show that results calculated using an $n$th-order GL expansion eventually improve with increasing $n$ for sufficiently small values of the order parameter. 

This paper is organized as follows. In Sects.~\ref{sec:GL_BCS} and \ref{sec:GL_NJL}, we derive the all-order formulas for the coefficients of the GL expansions and convergence radii in the BCS theory and NJL model, respectively. We discuss our results and give concluding remarks in Sect.~\ref{sec:discussion}. The technical details of the analysis are provided in appendices.

Throughout the paper, we use natural units $\hbar = c = k_B = 1$.

\section{GL expansion in BCS theory}
\label{sec:GL_BCS}

Let us first review the basics of BCS theory and then derive the formula that gives the GL coefficients to all orders. Although most of the results in this section are known in the literature (see, e.g.\ Ref.~\cite{ADG}), we review these for completeness and to discuss the similarities and differences with the results of the NJL model later. 

\subsection{Microscopic theory and free energy}

As we will not be interested in the electromagnetic responses in this paper, we can turn off the gauge fields. 
The BCS Lagrangian is then given by
\begin{equation}
    \label{BCS_lagrangian}
    \mathcal L_\text{BCS}
    = \psi^\dag 
    \left(
        i\partial_t + \frac{\bm \nabla^2} {2m} + \mu
    \right) \psi
    + \frac G2 (\psi^\dag \psi)^2\,.
\end{equation}
Here $\psi = (\psi_\uparrow, \psi_\downarrow)^\top$ is a two-component fermion spinor, $G$ is a four-fermion coupling constant used to model the attractive interaction between fermions, and $\mu$ is the chemical potential, which is equal to the Fermi energy $\epsilon_{\rm F} = k_{\rm F}^2 /(2m)$. Introducing the gap parameter $\Delta = {G}\langle \psi^\top C \psi \rangle/2$, which is assumed to be homogeneous for simplicity, with $C$ the charge conjugation matrix, one can show that the free energy (except for the terms that do not depend on $\Delta$) in the mean-field approximation is
\begin{equation}
    \label{F_BCS_schematic}
    F = \frac {\Delta^2} G - 
    2T \int_{|\bm k|\approx k_{\rm F}} 
    \frac{d^3k}{(2\pi)^3} 
    \ln \bigg[ 2 \cosh \left( \frac\beta2 \sqrt{\Delta ^2 + \xi_{\bm k}^2}\right) \bigg]\,,
\end{equation}
where $\beta$ is the inverse temperature $\beta=1/T$ and $\xi_{\bm k}=\epsilon_{\bm k}-\mu$ with $\epsilon_{\rm F} = |{\bm k}|^2/(2m)$. 
Here and below, we take $\Delta$ as real without loss of generality.

The integral should be taken near the Fermi surface over the region $k_{\rm F} - k_{\rm D} < |\bm k| < k_{\rm F} + k_{\rm D}$, where $k_{\rm D}$ is the Debye wavelength, which functions here as an ultraviolet (UV) cutoff. Near the Fermi surface, we can approximate $d^3k\approx 4\pi k_{\rm F}^2 d k$ and $\xi_{\bm k} \approx v_{\rm F}(k - k_{\rm F})$ with $v_{\rm F} = k_{\rm F}/m$ being the Fermi velocity. Therefore, assuming $k_{\rm D} \ll k_{\rm F}$, we have 
\begin{equation}
    \label{F_BCS}
    F = \frac {\Delta^2} G - 
    4 \rho T \int_0^{\omega_{\rm D}} d\xi \,
    \ln \bigg[ 2 \cosh \left( \frac\beta2 \sqrt{\Delta^2 + \xi^2} \right) \bigg]\,,
\end{equation}
where $\rho = k_{\rm F}^2 / (2 \pi^2 v_{\rm F})$ is the density of states per spin at the Fermi surface and $\omega_{\rm D} = v_{\rm F} k_{\rm D}$ is the Debye frequency. 

It is now clear that after an integral transformation $t = \beta \xi$, the integral in Eq.~(\ref{F_BCS}) is essentially $J_0(\beta \Delta, 0)$, except with the infinite upper limit replaced by a finite cutoff. Defining
\begin{align}
    \label{Jp_cut}
    J_{\ell}^\text{cut}(x, y; \lambda)
    &= \int_0^{\lambda} d t \,
    t^{\ell} \sum_{\zeta = \pm} \ln \bigg[ 2 \cosh
    \bigg(
        \frac
        {\sqrt{x^2 + t^2} + \zeta y} 2 
    \bigg) \bigg]\,,
\end{align}
we have
\begin{equation}
    \label{F_BCS_J0}
    F = \frac {\Delta^2} G - 
    2\rho T^2 J_0^\text{cut} 
    (\beta \Delta, 0; \beta \omega_{\rm D})\,.
\end{equation}
Notice that $\mu$ enters into the expression through the density of states $\rho$ in Eq.~(\ref{F_BCS})---hence, $\mu$ does not enter into the position of $y$ in $J_0(x, y)$, unlike the case of the dynamical quark mass in the NJL model that we study in the next section.

\subsection{nth-order coefficient formulas}
The GL expansion is a systematic expansion of the free energy $F$ in powers of the order parameter $\Delta$, in this case given by
\begin{equation}
    F = \alpha_2 \Delta^2
    + \alpha_4 \Delta^4
    + \cdots \; .
\end{equation}
Historically, GL theory was proposed as a phenomenological model of superconductivity, and the coefficients $\alpha_2$ and $\alpha_4$, often denoted $\alpha$ and $\beta$ (not to be confused with the inverse temperature) in the literature, were certain parameters. 
In the context of BCS theory, however, the coefficients---not only at the second and fourth orders~\cite{Gorkov1959}, but at all orders---are determined by the microscopic theory defined in Eq.~(\ref{BCS_lagrangian}). From Eq.~(\ref{F_BCS}) we see that the GL coefficients depend on the expansion of $J_0^\text{cut}(x, 0; \lambda)$ in powers of $x$. If we can find a general $n$th-order formula for these coefficients, then all the GL coefficients will immediately be known. Thus, our task is to calculate the coefficients appearing in
\begin{equation}
    \label{J0_cut_expansion}
    J_0^\text{cut}(x, 0; \lambda) 
    = c_2 x^2 + c_4 x^4 + \cdots \; ,
\end{equation}
via
\begin{equation}
c_{2n} = \frac{1}{n!}\partial_{x^2}^{n}  
    J_0^\text{cut}(x, 0; \lambda) \Big|_{x = 0}\,.
\end{equation}
Note that in every expansion considered here, we ignore the constant term $c_0$, because the physics is unchanged by an overall shift in the free energy. 

The trick to finding the (approximate) coefficients in Eq.~(\ref{J0_cut_expansion}) is to take $\lambda \to \infty$ on $c_{2n}$ that remain finite in this limit, which turn out to be those with $2n \geq 4$. In the weak-coupling limit, defined by $\rho G \ll 1$, $\omega_{\rm D}$ is large compared to other quantities characterizing the system, so taking $\lambda \to \infty$ is physically justified. Indeed, from the condition $\partial F/\partial \Delta = 0$ for Eq.~(\ref{F_BCS}) in the limit $T \rightarrow 0$, one easily finds that the zero-temperature gap $\Delta_0$ is given by
\begin{equation}
    \label{Delta_0}
    \Delta_0 = 2 \omega_{\rm D} e^{-1/(\rho G)}\,.
\end{equation}
Therefore, since in general $\Delta \leq \Delta_0$, we have $\lambda = \beta\omega_{\rm D} \gg \beta\Delta = x$, so that $\lambda$ is large relative to the other variables in Eq.~(\ref{Jp_cut}). We also comment later on how this approximation affects the radius of convergence. In this context, we could interpret $J_{\ell}(x, y)$ not as divergent and requiring regularization, but as an abbreviated form of a more complicated expression in which the infinite upper limit applies only to certain terms.

Taking a single derivative with respect to $x^2$ in Eq.~(\ref{Jp_cut}), letting $\ell = 0$ and $y = 0$, and converting the integrand to a Matsubara sum gives
\begin{align}
   \partial_{x^2}
    J_0^\text{cut}
    &= 
    2 \int_0^{\lambda} d t \,
    \frac{ \tanh 
    \left(\frac{1}{2} \sqrt{x^2 + t^2} \, \right)} 
    {4 \sqrt{x^2 + t^2}} \nonumber \\
    &= 
    2 \int_0^{\lambda} d t \,
    \sum_{k = 0}^\infty \frac1{\bar \omega_k^2 + x^2 + t^2}\,,
\label{J_0_derivative}
\end{align}
where $\bar \omega_k \equiv \beta \omega_k$ with $\omega_k= (2k + 1) \pi T$ being the fermionic Matsubara frequencies. 
Using the series expansion in $x^2$ of the summand, taking $(n-1)$ more derivatives with respect to $x^2$ under the integral and sum, and evaluating at $x = 0$, we find
\begin{align}
    \partial_{x^2}^{n} 
    J_0^\text{cut} \Big|_{x = 0} 
    &= 2 \int_0^{\lambda} d t \,
    \sum_{k = 0}^\infty \frac{(-1)^{(n-1)} (n-1)!}
    {(\bar \omega_k^2 + t^2)^{n}}\,.
    \label{J0_derivative_j+1}
\end{align}
This expansion of the summand in $x^2$ has a positive and finite radius of convergence for each choice of $\bar \omega_k^2 + t^2$, the smallest being $\bar \omega_1^2= \pi^2$.%
\footnote{In the bosonic case, on the other hand, the smallest value of $\omega_k^2$ is zero. Integrating this term and expanding in $x$ gives a series with an odd term $|x|^3$, which is not analytic in $x^2$ (see, e.g.\ Ref.~\cite{Laine2016}). Such a nonanalytic term does not appear in the fermionic case treated here due to nonzero $\omega_k$, which acts as an infrared (IR) cutoff.}
Since we will eventually substitute $x = \beta \Delta$, the convergence radius $x^2 = \pi^2$ corresponds to $\Delta = \pi T$. This foreshadows---but does not immediately imply---the final result, that the radius of convergence of the GL expansion is $\pi T$. 

The expression in Eq.~(\ref{J0_derivative_j+1}) is finite in the limit $\lambda \to \infty$ for $n \geq 2$. For these $n$, we can interchange the sum and integral, and then use the formula 
\begin{equation}
    \int_0^\infty \frac {d t} {(1 + t^2)^n}
    = \frac {(2n - 3)!!} {(2n - 2)!!} \frac\pi2\,,
    \label{integral_formula_1}
\end{equation}
which follows from a recursive relation that can be obtained using integration by parts. The remaining Matsubara sum can then be related to the Riemann zeta function, yielding
\begin{equation}
    \label{J0_cut_coef_formula}
    c_{2n\geq 4} = \frac {(-1)^{n + 1}} n
    \frac {(2n - 3)!!} {(2n - 2)!!} 
    \frac 1 {\pi^{2n - 2}}
    \big(1 - 2^{- (2n - 1)} \big) \zeta(2n - 1)
    \; .
\end{equation}
It follows from Eq.~(\ref{F_BCS_J0}) that the GL coefficients of order $2n \geq 4$ are given by
\begin{equation}
\alpha_{2n\geq 4} = - \frac{2\rho}{T^{2n - 2}}
c_{2n}
\,.
\end{equation}
In particular, this reproduces the well-known result for the GL coefficient of the $\Delta^4$ term~\cite{Gorkov1959}:
\begin{equation}
    \alpha_4 = \frac{7 \zeta(3)}{16} 
    \frac {\rho}{(\pi T)^2}\,.
\end{equation}

The result of the GL coefficient $c_2$ is also known~\cite{Gorkov1959}, but we also provide its derivation to make the paper self-contained.
For the coefficient $c_2$, we must proceed more carefully, because we cannot simply take $\lambda \to \infty$ (physically, $\omega_{\rm D}$ functions as a necessary UV cutoff). Using the expression in the first line of Eq.~(\ref{J_0_derivative}) at $x=0$, and then performing the integral gives
\begin{align}
    \partial_{x^2}  
    J^\text{cut}_0 \Big|_{x = 0}
    &= \frac12
    \left[
        \ln \left( \frac \lambda 2 \right) \tanh 
        \left( \frac \lambda 2 \right)
        - \int_0^{\lambda / 2} d t \, \ln(t) \operatorname{sech}^2(t)
    \right].
\end{align}
Now we clearly see a logarithmic UV divergence in the first term, although we can still take $\lambda \to \infty$ in the argument of $\tanh$ and on the remaining integral, the latter of which leads to
\begin{align}
\int_0^{\infty} d t \, \ln(t) \operatorname{sech}^2(t) = - \ln \left(\frac{4 e^{\gamma}}{\pi}\right)\,,
\end{align}
where $\gamma$ is the Euler--Mascheroni constant.
Thus we have
\begin{equation}
    \label{a_2_J0_cut}
    c_2 = \frac12 \ln
    \left( \frac{2 e^\gamma} \pi \lambda \right),
\end{equation}  
and the related result for BCS theory using Eq.~(\ref{F_BCS_J0}) is
\begin{equation}
    \label{alpha_BCS_omega_D}
    \alpha_2
    = \frac1G - \rho \ln
    \left( \frac{2 e^\gamma} \pi 
    \frac{\omega_{\rm D}} T \right)\,.
\end{equation}
The above result can be re-expressed in terms of the gap $\Delta_0$ at $T = 0$, using Eq.~(\ref{Delta_0}), giving
\begin{equation}
    \label{alpha_BCS_Delta_0}
    \alpha_2
    = \rho \ln \left(\frac{T}{T_{\rm c}} \right)\,,
\end{equation}
where $T_{\rm c} = e^\gamma \Delta_0 / \pi$ is the critical temperature of the superconducting state, at which $\alpha_2$ changes sign. Combining Eq.~(\ref{Delta_0}) with the formula for $T_{\rm c}$, we also find
\begin{equation}
    \label{omega_D_over_T_c}
    \frac{\omega_{\rm D}}{T_{\rm c}} =\frac\pi2 {e}^{-\gamma + 1/(\rho G)}\,.
\end{equation}

\subsection{Radius of convergence}
\label{subsec:BCS_Rad_Conv}

The radius of convergence of the expansion in Eq.~(\ref{J0_cut_expansion}) can essentially be read off of the coefficient formula (\ref{J0_cut_coef_formula}). We have $(x^2)_\text{conv} = \pi^2$ when viewed as a series in $x^2$, or equivalently, $x_\text{conv} = \pi$ when viewed as a series in $x$. To make the argument rigorous, one can apply the ratio test, $(x^2)_\text{conv} = \lim_{n\to\infty} |c_{2n + 2} / c_{2n}|$. Finally, Eq.~(\ref{F_BCS_J0}) shows that the GL expansion is essentially given by the same series, with $x = \beta \Delta$, so we have $\Delta_\text{conv} = x_\text{conv} T = \pi T$.

Let us explain the exact meaning and implications of this result. The radius of convergence $\pi$ applies technically to the series whose coefficients $c_{2n}$ are given by Eq.~(\ref{J0_cut_coef_formula}), but these are computed in the limit $\lambda \to \infty$, in which the original quantity $J^\text{cut}_0$ becomes ill-defined.  Nonetheless, it is easy to see that the true coefficients in the expansion of $J_0^\text{cut}$, keeping $\lambda$ finite, are bounded from above in magnitude by $c_{2n}$ in Eq.~(\ref{J0_cut_coef_formula}). This follows because the integrand of Eq.~(\ref{J0_derivative_j+1}) is either entirely positive or entirely negative, so increasing the upper limit of integration must strictly increase its absolute value. Since $c_{2n}$ of Eq.~(\ref{J0_cut_coef_formula}) lead to a series with radius of convergence $\pi$, and the series for $J_0^\text{cut}$ with $\lambda$ finite has its coefficients bounded by the former, we can actually conclude that the radius of convergence is {\it at least} $\pi$ for the true expansion of $J_0^\text{cut}$, and hence the GL expansion has radius of convergence of at least $\pi T$.

When using the GL expansion to approximate and solve the gap equation, the key question is whether the true gap $\Delta_\text{exact}$ falls within this convergence radius, $\Delta_\text{exact} < \Delta_\text{conv}$. When this condition is satisfied, then the solution of the $N$th-order GL expansion, $\Delta_{\text{GL},N}$, will converge to $\Delta_\text{exact}$ as $N \to \infty$. This is the convergence of the GL \textit{solution}. Note that this should be distinguished from the convergence of the GL \textit{expansion} itself over some (possibly vanishing) range $0 \leq \Delta < \Delta_\text{conv}$ for each fixed $T$.

\begin{figure}
    \centering
    \subfloat{\includegraphics[width=.47 \textwidth]{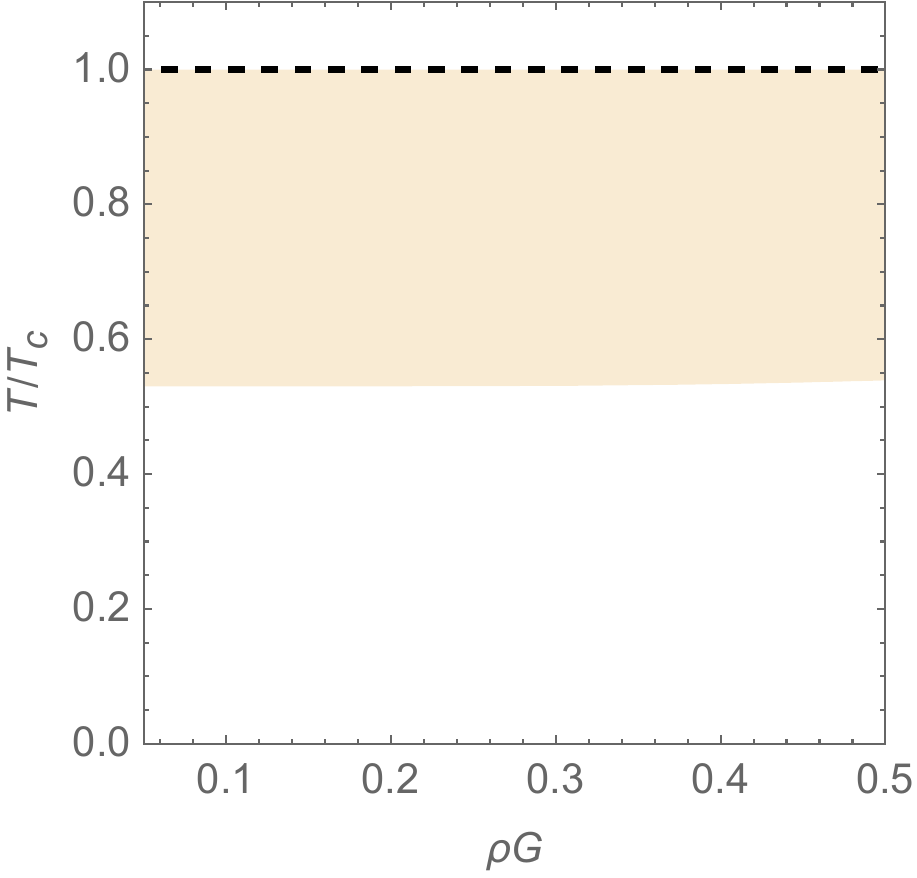}} \hfill
    \subfloat{\includegraphics[width=.45 \textwidth]{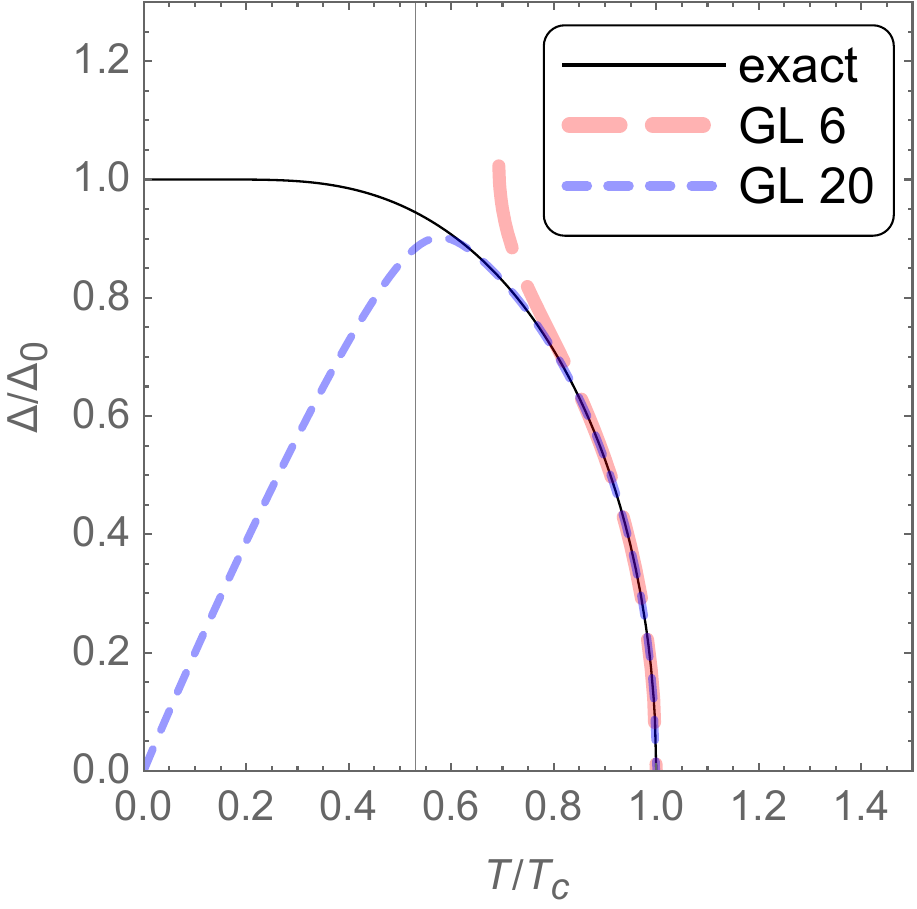}} \hfill
    \caption{{Left:} Region in which the GL solution converges for BCS theory. The black dashed line denotes the second-order phase transition, and the orange shaded region indicates where the GL solution converges with $\Delta > 0$. {Right:} BCS gap versus temperature, computed by solving the gap equation using the exact free energy (black), a 6th-order GL expansion (pink, dashed), and a 20th-order GL expansion (blue, dotted). GL solution should converge to the right of the vertical gray line, i.e.\ $T/T_{\rm c} \geq 0.53$. These data are computed for $\rho G = 0.18$.}
    \label{fig:BCS_conv_radius}
\end{figure}

The range of temperatures over which the GL solution converges is indicated in the left panel of Fig.~\ref{fig:BCS_conv_radius} (orange shaded region). We searched numerically for the temperature at which $\Delta_\text{exact}$ drops below $\Delta_\text{conv}$ over a range of parameter values for $\rho G$. For each $\rho G$, the value of $\omega_{\rm D}/T_{\rm c}$ is fixed by Eq.~(\ref{omega_D_over_T_c}), so $\rho G$ is the only parameter of the theory when all quantities are expressed relative to $T_{\rm c}$. Note that the lower boundary of the convergence region is nearly constant over the given range of $\rho G$ (e.g.\ for $\rho G \leq 0.21$ the GL solution converges when $T/T_{\rm c} \geq 0.53$). The right panel of Fig.~\ref{fig:BCS_conv_radius} shows $\Delta$ vs.\ $T$ computed from the gap equation and also from 6th- and 20th-order GL expansions. Both GL expansions are reliable near $T_{\rm c}$, as expected, and the 20th-order GL solutions remain reliable over a wider range of $T$. 

Let us also comment on how the GL solutions in the right panel of Fig.~\ref{fig:BCS_conv_radius} are calculated. Since solving the stationary equation $\partial F/\partial \Delta = 0$ for an $N$th-order GL expansion involves finding the roots of an $(N - 1)$th-order polynomial in $\Delta$, one should expect in general to find up to $N - 1$ potential solutions. Moreover, if the largest coefficient in such an $N$th-order GL expansion is negative, then the free energy must eventually decrease toward $-\infty$ as $\Delta \to \infty$, so one cannot determine the GL solutions simply by finding the global minimum. Instead, we essentially just search for the local minimum with the smallest free energy over $0 \leq \Delta \leq \Delta_\text{conv}$. Although many more real roots of the stationary equation can appear with increasing $N$, in practice these occur far outside the radius of convergence, and hence they are easily neglected. The only additional complication is that the minimum in the 20th-order case gradually shifts and moves outside the convergence region as $T/T_{\rm c}$ drops below $0.53$, and we continue to plot this minimum in the right panel of Fig.~\ref{fig:BCS_conv_radius} (blue dotted curve left of vertical line). We treat this as the solution because it remains close to the convergence region and is continuously connected to the solution at higher $T$. Note that this minimum remains well defined at all $T$ because  $\alpha_{20}$ is positive; by contrast, the red dashed curve vanishes for $T/T_{\rm c} < 0.69$ because the relevant local minimum at 6th order truly vanishes at these temperatures.

\section{GL expansion in the NJL Model}
\label{sec:GL_NJL}

In this section, we show how the expression $J_2(x, y)$ appears in the GL expansion for the dynamical quark mass in the NJL model and how it can be regularized in this setting. We then solve for the $n$th-order GL coefficients and convergence radius.

\subsection{Microscopic theory and free energy}

The two-flavor NJL Lagrangian is given by~\cite{Nambu:1961tp, Klevansky:1992qe, Hatsuda:1994pi, Buballa:2003qv}:
\begin{equation}
    \label{NJL_lagrangian}
    \mathcal L_\text{NJL} = \bar \psi (i\gamma^\mu \partial_\mu - \hat m) \psi + G \big[(\bar \psi \psi)^2 + (\bar \psi i\gamma^5 \tau^a \psi)^2 \big].
\end{equation}
Here $\psi = ({\rm u, d})^\top$ is a two-flavor quark field, $\hat m = \operatorname{diag}(m_{\rm u}, m_{d})$ is the bare quark mass matrix in flavor space, $G$ is a four-fermion coupling constant, and $\tau^a$ are the Pauli matrices acting in flavor space. This model retains the approximate chiral symmetry of QCD, which becomes exact in the chiral limit $\hat m \to 0$. We will assume the chiral limit from now on.

In vacuum, chiral symmetry is spontaneously broken by the formation of a quark--antiquark pairing $\sigma = \langle \bar \psi \psi \rangle$, called the chiral condensate. To study the behavior of the condensate $\sigma$ as a function of temperature $T$ and quark chemical potential $\mu$, we expand Eq.~(\ref{NJL_lagrangian}) about the ansatz: 
\begin{equation}
    \label{CC_ansatz}
    \langle \bar \psi \psi \rangle = \sigma, 
    \qquad 
    \langle \bar \psi i\gamma^5 \tau^a \psi \rangle = 0 \quad (a = 1,2,3).
\end{equation}
We note that at sufficiently large densities, or even small finite densities and in the presence of a magnetic field, this ansatz is known to be disfavored over various spatially inhomogeneous condensates within the mean-field approximation~\cite{Buballa:2014tba}, the simplest being the so-called chiral density wave, given by $\langle \bar \psi \psi \rangle + i\langle \bar \psi i\gamma^5 \tau^3 \psi \rangle = \sigma e^{iqz}$. In this paper, we restrict to the homogeneous case for simplicity.

In the mean-field approximation, one can show that the free energy is given by~\cite{Hatsuda:1994pi, Buballa:2003qv}:
\begin{equation}
    \label{omega_CC_pre_reg}
    \Omega_\text{pre-reg} 
    = \frac {M^2} {4G} - 2 N_{\rm f} N_{\rm c}
    \int
    \frac {d^3 k} {(2 \pi)^3}
    \left[
        E_{\bm k} + \sum_{\zeta = \pm 1} T \ln
        \left(
            1 + e^{-\beta (E_{\bm k} + \zeta \mu)}
        \right)
    \right],
\end{equation}
where $M = -2G \sigma$ can be interpreted as the dynamical mass of quarks due to the chiral condensate; $N_{\rm f}=2$ and $N_{\rm c}=3$ are the numbers of flavors and colors, respectively; and $E_{\bm k} = \sqrt{\bm k^2 + M^2}$. Making the transformations $x = \beta k$ and $y = \beta \mu$, we find
\begin{equation}
    \Omega_\text{pre-reg} 
    = \frac {M^2} {4G} - 
    \frac {N_{\rm f} N_{\rm c}} {\pi^2} T^4 
    J_2 (\beta M, \beta \mu)\,.
\end{equation}
Here, the divergence of $J_2$ reflects the zero-point vacuum energy, and this divergence must be regularized in order to carry out numerical calculations. Several schemes are in common use \cite{Klevansky:1992qe}. For consistency with the last section, we will impose a simple momentum cutoff $\Lambda$, and take $\Lambda \to \infty$ on any terms that remain finite in this limit. Such a cutoff clearly violates Lorentz invariance, which can lead to undesired consequences when the condensate under consideration is inhomogeneous \cite{Buballa:2014tba}. In that context, covariant regularization methods, such as the Schwinger proper time method  \cite{Schwinger:1951nm}, are typically used instead. Since we consider only homogeneous condensates in this paper, the following analytical results based on a simple cutoff are physically meaningful. However, the numerical accuracy of the approximation $\Lambda = \infty$ is somewhat limited in this setting, as we will discuss in Sect.~\ref{subsec:RoC_CC}. It is also possible to solve for the GL coefficients using the proper time method, and in fact, that allows for analytical formulas for the GL coefficients even without taking the limit $\Lambda \to \infty$; these results are given in Appendix \ref{appendix:cc_proper_time}. For the sake of accuracy, we use the proper time method in the numerical results presented in Fig.~\ref{fig:CC_schw_conv_radius}.

Imposing the momentum cutoff $|\bm k| < \Lambda$ in Eq. (\ref{omega_CC_pre_reg}), we find 
\begin{equation}
\label{omega_CC_cut}
    \Omega^\text{cut} 
    = \frac {M^2} {4G} - 
    \frac {N_{\rm f} N_{\rm c}} {\pi^2} T^4 
    J_2^\text{cut} (\beta M, \beta \mu; \beta \Lambda)\,.
\end{equation}

\subsection{nth-order coefficient formulas}
\label{subsec:CC_coef_formulas}

As in the previous case, the GL expansion takes the form
\begin{equation}
    \Omega = \alpha_2 M^2
    + \alpha_4 M^4 + \cdots .
\end{equation}
The coefficient $\alpha_4$ was previously computed in Ref.~\cite{Yamamoto:2007ah}, but the expression for the generic coefficient $\alpha_{2n}$ has not been calculated explicitly, to the best of our knowledge. These coefficients are determined by the expansion coefficients of $J_2^\text{cut}(x, y; \lambda)$ in powers of $x$,
\begin{equation}
    \label{J2_cut_series}
    J_2^\text{cut}(x, y; \lambda)
    = c_2 x^2 + c_4 x^4 + \cdots,
\end{equation}
via
\begin{equation}
c_{2n} = \frac{1}{n!}\partial_{x^2}^{n}  
    J_2^\text{cut}(x, y; \lambda) \Big|_{x = 0}\,,
\end{equation}
which we will now derive. Finally, once the $c_{2n}$ coefficients are known, the $\alpha_{2n}$ coefficients follow immediately from Eq. (\ref{omega_CC_cut}),
\begin{equation}
\label{alpha_vs_c_CC}
    \alpha_{2n}
    = \frac{\delta_{n1}}{4G}
    - \frac{N_\text f N_\text c}{\pi^2}
    \frac{c_{2n}}{T^{2n - 4}}\Big|_{y = \beta\mu, \lambda = \beta\Lambda}\,.
\end{equation}

The two differences from the BCS case are that we now have $J_2$ rather than $J_0$, and we now have $y \neq 0$. The appearance of $J_2$ adds a factor of $t^2$ to the integrand, which strengthens the UV divergence, and as a result, we will need to keep $\lambda$ finite on the $c_4$ coefficient in addition to $c_2$. Having $y\neq 0$ complicates the algebra, but does not significantly affect the overall approach.

Following the same procedure as in the BCS case, we start by taking a single derivative with respect to $x^2$, giving
\begin{align}
   \partial_{x^2}
    J_2^\text{cut}
    &= 
    \int_0^{\lambda} d t 
    \sum_{\zeta = \pm 1}
    \frac{t^2}{4 \sqrt{x^2 + t^2}} 
    \tanh \left(
    \frac{\sqrt{x^2 + t^2} + \zeta y}{2} 
    \, \right) 
    \nonumber \\
    &= 
    \int_0^{\lambda} d t \,
    \sum_{\zeta = \pm 1}
    \frac{t^2}{\sqrt{x^2 + t^2}} 
    \sum_{k = 0}^\infty 
    \frac{\sqrt{x^2 + t^2} + \zeta y}
    {\bar \omega_k^2 
    + (\sqrt{x^2 + t^2} + \zeta y)^2} \, ,
\end{align}
where $\bar \omega_k = (2k + 1)\pi$, as before. After applying the identity
\begin{equation}
    \sum_{\zeta = \pm 1}
    \frac1b \cdot 
    \frac{b + \zeta c}{a^2 + (b + \zeta c)^2}
    = \sum_{\zeta = \pm 1}
    \frac1{b^2 + (a + i\zeta c)^2}\,,
\end{equation}
we obtain
\begin{equation}
    \label{d_J2_cut}
    \partial_{x^2} J_2^\text{cut}
    = 2 \operatorname{Re}
    \sum_{k = 0}^\infty 
    \int_0^\lambda d t
    \frac{t^2}{x^2 + t^2 + (\bar \omega_k + iy)^2}\,.
\end{equation}
At this point, in close analogy to the comment after Eq.~(\ref{J0_derivative_j+1}), we may see a hint of the final result for the radius of convergence. If one expands the integrand in powers of $x^2$, the resulting series has radius of convergence $x_\text{conv}^2 = |t^2 + (\bar{\omega}_k + iy)^2|$. Taking $t \to 0$ then gives $x_\text{conv}^2 = |\bar{\omega}_k + iy|^2$, the minimum of which is $x_\text{conv}^2 = \pi^2 + y^2$; recalling that $x = \beta M$ and $y = \beta \mu$, this corresponds to $M_\text{conv} = \sqrt{\mu^2 + (\pi T)^2}$. However, this heuristic argument is fairly crude, partly because the quantity $|t^2 + (\bar{\omega}_k + iy)^2|$ need not be minimized in the limit $t \to 0$. 
We therefore proceed to the rigorous proof.

Taking $n - 1$ more derivatives of Eq.~(\ref{d_J2_cut}) with respect to $x^2$, introducing a variable $s = t/(\bar \omega_k + iy)$, and letting $\lambda \to \infty$, we find
\begin{equation}
    \label{d^n_J2cut}
    \partial_{x^2}^n J_2^\text{cut} \Big|_{x = 0}
    = 2 (-1)^{n - 1} (n - 1)! \operatorname{Re}
    \sum_{k = 0}^\infty 
    \frac 1 {(\bar \omega_k + iy)^{2n - 3}}
    \int_0^\infty d s
    \frac{s^2}{(s^2 + 1)^n}\,.
\end{equation}
The integral is easily performed by using Eq. (\ref{integral_formula_1}), giving
\begin{equation}
    \int_0^\infty d s 
    \frac{s^2} {(1 + s^2)^n}
    = \frac {(2n - 5)!!} {(2n - 2)!!} \frac\pi2 \, .
    \label{integral_formula_3}
\end{equation}
Notice that the sum in Eq.~(\ref{d^n_J2cut}) converges when $n \geq 3$. Identifying the sum with the standard series representation of the $n$th-order polygamma function $\psi^{(n)}(z) = (-1)^{n+1} n! \sum_{k=0}^{\infty} 1/(z+k)^{n+1}$, we finally find
\begin{equation}
\label{c2n}
    c_{2n\geq 6} = 
    \frac{(-1)^n}
    {n! 2^n (2\pi)^{2n - 4} (2n - 4)!!}
    \operatorname{Re} \psi^{(2n - 4)}
    \left( \frac12 + i\frac y{2\pi} \right)\,.
\end{equation}

For $c_2$ and $c_4$, it is more convenient to avoid writing $J_2$ as a Matsubara sum, starting instead from the expression
\begin{align}
\label{J2_cut}
    J_2^\text{cut}
    &= 
    \int_0^{\lambda} d t \, t^2
    \left[
    \sqrt{x^2 + t^2}
    + \sum_{\zeta = \pm 1}
    \ln 
    \left(
    1 + e^{-\sqrt{x^2 + t^2} + \zeta y}
    \right)
    \right].
\end{align}
For $c_2$, it is straightforward to calculate
\begin{align}
    \partial_{x^2} J_2^\text{cut} 
    \Big|_{x = 0}
    &= \frac{\lambda^2} 4 +
    \frac12 \sum_{\zeta = \pm 1}
    \left[
        \lambda \ln 
        \left(
        1+e^{-\lambda + \zeta y}
        \right)
        - \operatorname{Li}_2
        \left(
            -e^{-\lambda + \zeta y}
        \right)
        + \operatorname{Li}_2
        \left(
            -e^{\zeta y}
        \right)
    \right],
    \label{derivative_CC_a2_med}
\end{align}
where $\Li_n(x) = \sum_{k=1}^{\infty} x^k/k^n$ is the $n$th polylogarithm. The first two terms under the sum in Eq. (\ref{derivative_CC_a2_med}) vanish as $\lambda \to \infty$, and for the last term we can apply the identity \cite{Apostol1976}, 
\begin{equation}
    \Li_n(-e^{y})+(-1)^n\Li_n(-e^{-y}) = -\frac{(2\pi i)^n}{n!} B_n\left(\frac12+\frac y{2\pi i}\right),
\end{equation}
where $B_n(x)$ is the $n$th Bernoulli polynomial. The result is 
\begin{equation}
\label{c2}
    c_2 = \frac14\left(\lambda^2 - y^2 - \frac{\pi^2}3 \right)\,.
\end{equation}

The $c_4$ coefficient is more subtle because a naive separation of Eq. (\ref{J2_cut}) into vacuum and medium contributions leads to IR divergences. An efficient way to compute this coefficient is to note that after commuting the derivatives $\partial_{x^2}^2$ with $t^2$ in the integrand, we can transform them as $\partial_{x^2} \to (2t)^{-1} \partial_t$. This allows us to take the limit $x \to 0$ and then take derivatives with respect to $t$. Performing only the rightmost derivative, we find
\begin{equation}
    \partial_{x^2}^2 J_2^\text{cut} 
    \Big|_{x = 0}
    = \frac14 \int_0^\lambda d t \,
    t \, \partial_t \frac1t
    \left(
    1 - \sum_{\zeta = \pm 1} \frac1 {1 + e^{t + \zeta y}}
    \right).
\end{equation}
Integrating by parts, taking $\lambda \to \infty$ on the boundary terms, and by applying the formula \cite{Gyory},
\begin{equation}
    \lim_{\lambda \to \infty} 
    \left[
    \ln \left(\frac{\lambda}{2\pi} \right) - 
    \int_0^\lambda \frac{d t}t
    \left(
    1 - \sum_{\zeta = \pm 1} \frac1{1 + e^{t + \zeta y}}
    \right)
    \right]
    = \operatorname{Re} \psi \left( \frac12 + i\frac y{2\pi} \right)\,,
\end{equation}
where $\psi(z)$ is the digamma function defined by $\psi(z)={d} \ln \Gamma(z)/d z$ with $\Gamma(z)$ the gamma function, we obtain
\begin{equation}
    \partial_{x^2}^2 J_2^\text{cut} \Big|_{x = 0}
    \approx \frac14 
    \left[
    1 - \ln \left(\frac{\lambda}{2\pi} \right) 
    + \operatorname{Re} \psi \left( \frac12 + i\frac y{2\pi} \right)
    \right]\,,
\end{equation}
for large $\lambda$, and accordingly,
\begin{align}
\label{c4}
    c_4 &= \frac18 
    \left[
    1 - \ln \left(\frac{\lambda}{2\pi} \right) 
    + \operatorname{Re} \psi \left( \frac12 + i\frac y{2\pi}         \right)
    \right]\,.
\end{align}

Finally, applying Eq. (\ref{alpha_vs_c_CC}), we have
\begin{align}
    \alpha_2 
    &= 
    \frac1{4G} - \frac{N_\text f N_\text c}{\pi^2}
    \frac14\left(\Lambda^2 - \mu^2 - \frac{\pi^2}3 T^2 \right)\,,
    \label{alpha_2}
    \\
    \alpha_4 
    &= 
    - \frac{N_\text f N_\text c}{\pi^2}
    \frac18 
    \left[
    1 - \ln \left(\frac{\Lambda}{2\pi T} \right) 
    + \operatorname{Re} \psi 
    \left( \frac12 + i\frac \mu{2\pi T} \right)
    \right]\,,
    \\
    \alpha_{2n\geq 6} 
    &=
    - \frac{N_\text f N_\text c}{\pi^2}
    \frac{(-1)^n}
    {n! 2^n (2\pi T)^{2n - 4} (2n - 4)!!}
    \operatorname{Re} \psi^{(2n - 4)}
    \left( \frac12 + i\frac \mu{2\pi T} \right)\,.
    \label{alpha_2n}
\end{align}

\subsection{Radius of convergence}
\label{subsec:RoC_CC}

We will show that the expansion of $J_2^\text{cut}(x, y; \lambda)$ has radius of convergence $x_\text{conv} = \sqrt{\pi^2 + y^2}$. Then, from Eq.~(\ref{omega_CC_cut}), it will follow that the GL expansion has convergence radius $M_\text{conv} = (x_\text{conv}|_{y = \beta \mu}) T = \sqrt{\mu^2 + (\pi T)^2}$.

First, since the convergence of any series only depends on its infinite tail, and not on any finite initial segment, we can focus on the coefficients $c_{2n \geq 6}$ given by Eq.~(\ref{c2n}). It is straightforward to show that the radius of convergence is {\it at least} $\sqrt{\pi^2 + y^2}$. Using the inequality $|\operatorname{Re} (z)| \leq |z|$ for complex $z$, we have
\begin{equation}
    \label{a_2n_med_inequality}
    \big| c_{2n} \big|
    \leq \frac 1 {n! \, 2^{n} (2n - 4)!!} 
    \frac 1 {(2 \pi)^{2n - 4}}
    \bigg|
    \psi^{(2n - 4)}\left(
    \frac12 + i\frac y {2 \pi}
    \right) \bigg| \eqqcolon b_{2n} \,,
\end{equation}
for $2n \geq 6$.

Using the identity $\psi^{(n)}(x) = (-1)^{n+1} n! \, \zeta(n + 1, x)$, where $\zeta(s,a)=\sum_{k=0}^{\infty}1/(k+a)^{s}$ is the Hurwitz zeta function, one can show that
\begin{equation}
    \frac {\psi^{(n)}(x)} {\psi^{(n + 2)}(x)}
    \xrightarrow{n \to \infty} \frac 1 {(n + 1) (n + 2)} x^2
\end{equation}
for $\operatorname{Re}(x) > 0$, in the sense that the ratio of the quantities on the left and right sides of the arrow approaches unity as $n \to \infty$. It follows that 
\begin{align}
    \lim_{n \to \infty}
    \Bigg| \frac {b_{2n}} 
    {b_{2n + 2}} \Bigg|
    &= \pi^2 + y^2.
\end{align}
Thus, the series for $J_2^\text{cut}(x, y; \lambda)$ in Eq.~(\ref{J2_cut_series}) is bounded in magnitude by a series with radius of convergence $x_\text{conv} = \sqrt{\pi^2 + y^2}$, so the original series has a convergence radius of {\it at least} this value. In fact, one can show the exact convergence radius is indeed $\sqrt{\pi^2 + y^2}$, but the proof is more involved. In particular, if we do not take the upper bound of the coefficients using $|\operatorname{Re} (z)| \leq |z|$, then the ratio test does not work, because the quantity $[\operatorname{Re} \psi^{(2n - 4)}(\frac12 + i\frac{y}{2\pi})] \allowbreak / [\operatorname{Re} \psi^{(2n - 2)}(\frac12 + i\frac{y}{2\pi})]$ oscillates erratically. One can instead use the Cauchy--Hadamard theorem; we sketch this proof in Appendix~\ref{appendix:Cauchy-Hadamard}.

Although the results of this section were computed using a simple momentum cutoff $\Lambda$, the radius of convergence is unaffected when using Schwinger proper time regularization. Whereas the former cutoff scheme involves taking the limit $\Lambda \rightarrow \infty$ to obtain parts of $c_{2n}$ and $\alpha_{2n}$ analytically, the latter scheme allows deriving their analytic expressions even without taking such a limit for a cutoff parameter $\Lambda$. As shown in Appendix \ref{appendix:cc_proper_time}, the $\alpha_{2n \geq 6}$ coefficients are almost the same as in Eq.~(\ref{alpha_2n}), except with small corrections proportional to inverse powers of $\Lambda$. The key fact is that these correction terms decay faster than the $\alpha_{2n}$ coefficients calculated in this section, so they do not increase the convergence radius. We prove these statements in Appendix~\ref{appendix:cc_proper_time}.

Let us also mention that the assumption of $\Lambda = \infty$ made when deriving Eqs.~(\ref{alpha_2})--(\ref{alpha_2n}) is only a rough approximation in the NJL setting. 
The values for the two parameters $\Lambda$ and $G$ are determined by a choice of values for $M_0$ and $f_\pi$, where $M_0$ is the dynamical quark mass and $f_\pi$ is the pion decay constant at $T=\mu=0$. Choosing $M_0 = 300\,\text{MeV}$ and $f_\pi = 88\,\text{MeV}$ in the chiral limit~\cite{Nickel:2009wj, Buballa:2014tba}, we find $\Lambda = 614\,\text{MeV}$ and $G\Lambda^2 = 2.15$ \cite{Klevansky:1992qe}. 
With these values, one can numerically find $T_{\rm c} = 182\,\text{MeV}$ at $\mu = 0$, and $\mu_{\rm c} = 311\,\text{MeV}$, where $\mu_c$ is the chemical potential at which $M$ vanishes in a first-order transition with $T = 0$ fixed. Thus, $\Lambda$ is not especially large on the scale of other characteristic quantities of the system, so taking $\Lambda \to \infty$ is not fully justified. For instance, finding $T_{\rm c}$ by setting $\alpha_2 = 0$ in Eq.~(\ref{alpha_2}) gives $T_{\rm c} = 165\,\text{MeV}$. Although the series with coefficients given by Eqs.~(\ref{alpha_2})--(\ref{alpha_2n}) converges with radius $\sqrt{\mu^2 + (\pi T)^2}$, the value does not converge exactly to $\Omega$, even within this radius. 

\begin{figure}
    \centering
    \subfloat{\includegraphics[width=.45 \textwidth]{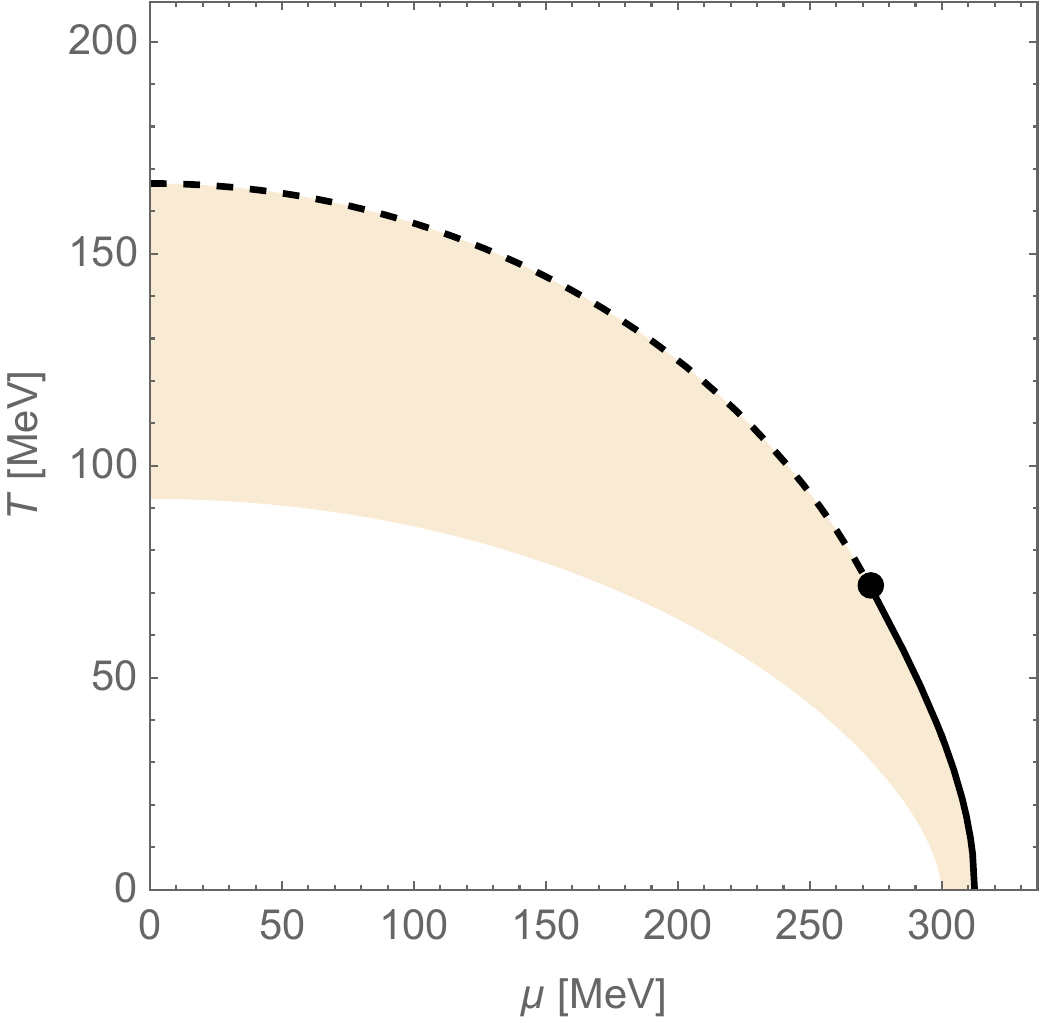}} \hfill
    \\
    \subfloat{\includegraphics[width=.45 \textwidth]{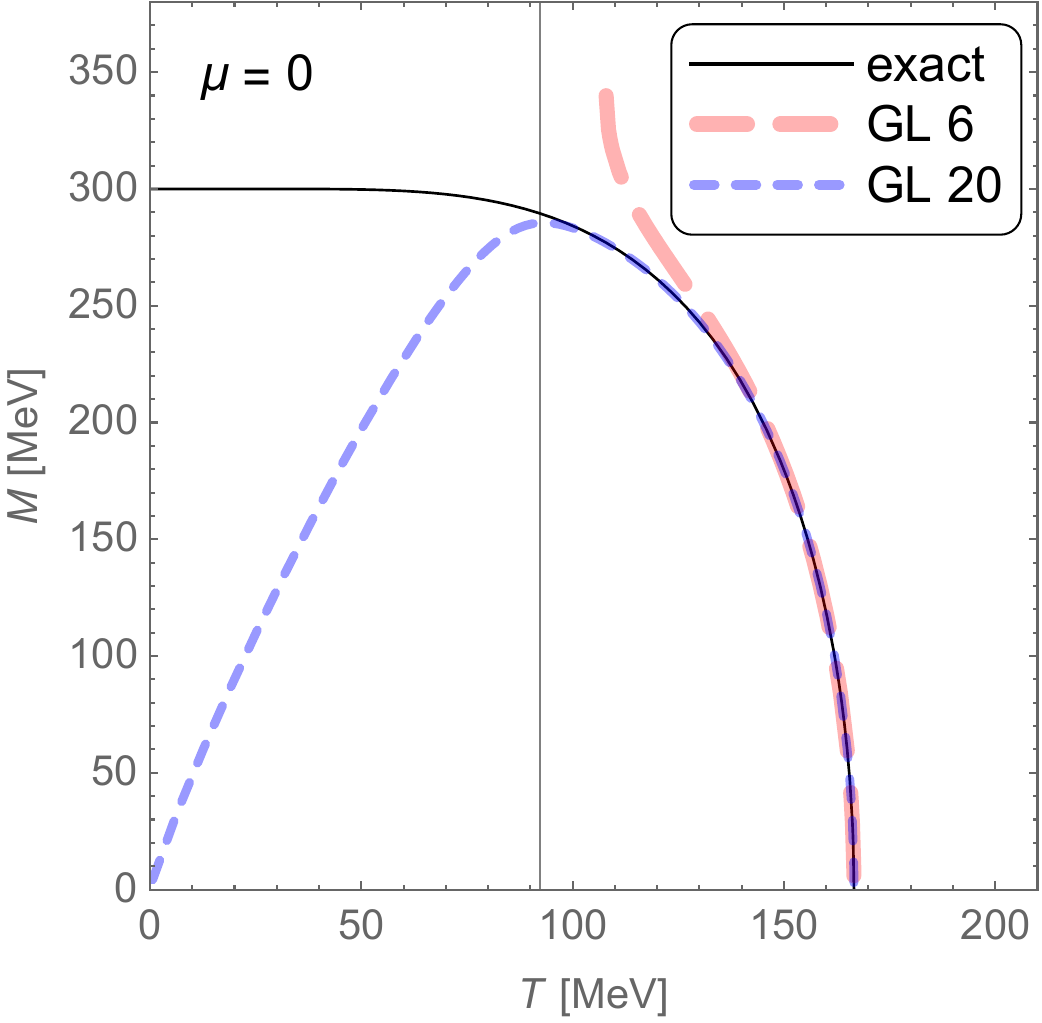}} \hfill
    \subfloat{\includegraphics[width=.46 \textwidth]{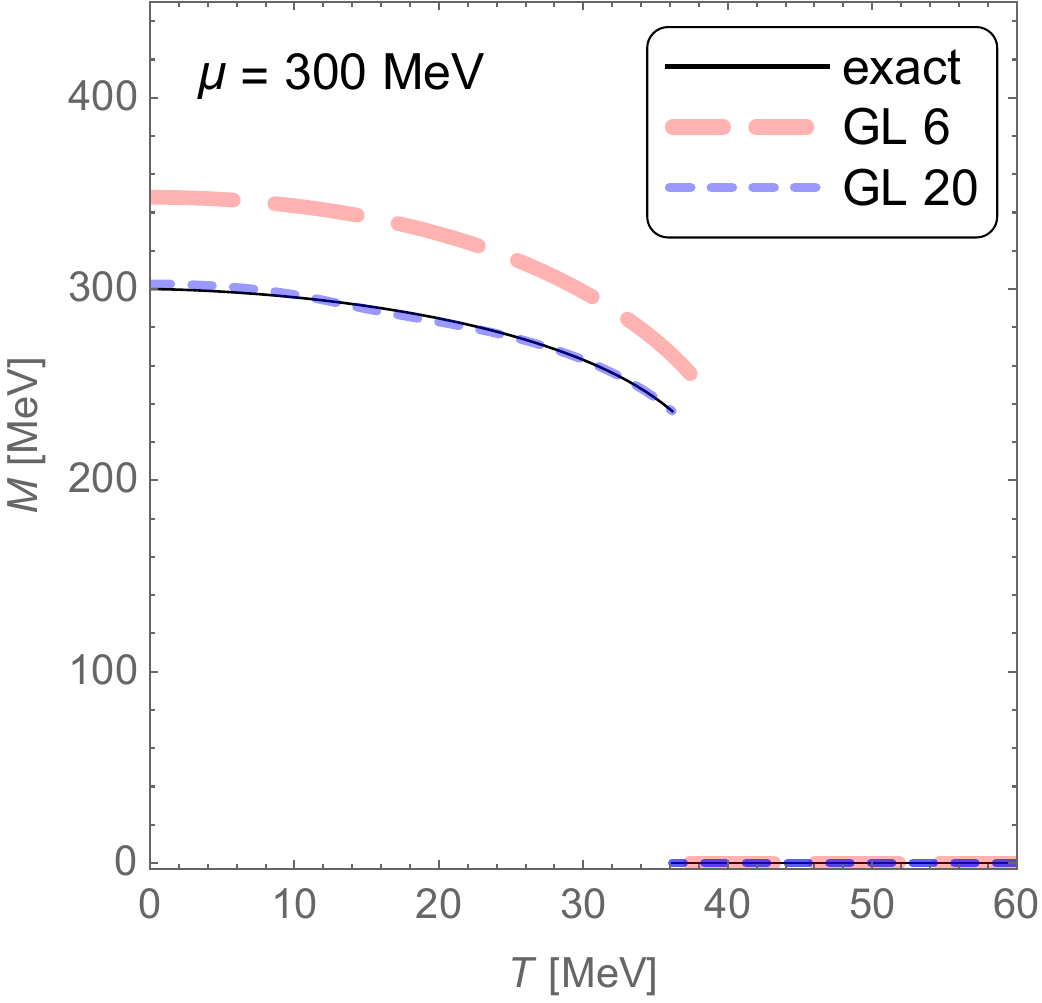}} \hfill
    \caption{{Top:} Region in which the GL solution for the dynamical quark mass converges in the NJL model using Schwinger proper time regularization. The black curve shows the critical temperature, with solid (dashed) lines indicating the first-order (second-order) phase transitions. The orange shaded region indicates where the GL solution converges when $M > 0$. 
    {Bottom:} Dynamical quark mass versus temperature at $\mu = 0$ (left) and $\mu = 300\,\text{MeV}$ (right), computed by solving the gap equation using the exact free energy (black), a 6th-order GL expansion (pink, dashed), and a 20th-order GL expansion (blue, dotted).
    }
    \label{fig:CC_schw_conv_radius}
\end{figure}

These issues are avoided when using Schwinger proper time regularization, described in Appendix~\ref{appendix:cc_proper_time}, because there the finite size of $\Lambda$ is captured in the analytical formulas for the coefficients. Therefore, the numerical results for this section, presented in Fig.~\ref{fig:CC_schw_conv_radius}, are computed using the Schwinger method. In this regularization scheme, again choosing $M_0 = 300\,\text{MeV}$ and $f_\pi = 88\,\text{MeV}$, we find $\Lambda = 634\,\text{MeV}$ and $G\Lambda^2 = 6.02$. We calculated $T_{\rm c}$ over the range $0 \leq \mu \leq 312\,\text{MeV}$, and we determined the region in the $\mu$-$T$ plane below $T_{\rm c}$ where $M < M_\text{conv}$, i.e.\ where the GL solution converges (top panel, orange shaded region). The bottom panels of Fig.~\ref{fig:CC_schw_conv_radius} show $M$ vs.\ $T$ computed from the gap equation and also from the GL expansion at orders 6 and 20. For $\mu = 0$ (bottom left panel), the GL solution converges only when $T > 92\,\text{MeV}$ (right of the vertical line); both the $6$th- and $20$th-order solutions are very accurate near $T_{\rm c}$, but the $6$th-order solution becomes noticeably inaccurate at $T \lesssim 130\,\text{MeV}$. On the other hand, when $\mu = 300\,\text{MeV}$ (bottom right panel), the phase transition is first order, and the $20$th-order solution is a noticeable improvement over the $6$th-order solution across the entire range $0 < T < T_{\rm c}$. Moreover, the GL solution converges over this entire range, and hence the $20$th-order solutions are very reliable all the way down to $T = 0$. This can be understood by considering the radius of convergence formula at $T = 0$, which reduces to $M_\text{conv} = \mu$, and noticing that $M \leq 300\,\text{MeV}$ for $\mu > 300\,\text{MeV}$.

We also note that the method used for determining the solutions of the 6th- and 20th-order GL expansions, plotted in the lower panels of Fig.~\ref{fig:CC_schw_conv_radius}, is exactly the same as in the BCS case, as discussed in the last paragraph of Sect.~\ref{subsec:BCS_Rad_Conv}.

As mentioned above and shown in Appendix \ref{appendix:cc_proper_time}, the radius of convergence when using the proper time method is also $\sqrt{\mu^2 + (\pi T)^2}$. One could also ask about the radius of convergence in the cutoff method if {\it not} using the approximation $\Lambda = \infty$, i.e.\ if instead one were to calculate the GL coefficients from Eq.~(\ref{omega_CC_cut}) numerically at finite $\Lambda$. It is possible to show that in this case the radius of convergence is still given by at least $\sqrt{\mu^2 + (\pi T)^2}$ when $M > 0$; we sketch a proof in Appendix \ref{appendix:RoC_cut_finite_lambda}.

\section{Discussions and outlook}
\label{sec:discussion}

In this paper, we studied the convergence of the GL expansion for two prominent examples---the BCS theory for superconductivity and the NJL model for chiral symmetry breaking at finite temperature $T$ and chemical potential $\mu$. We have shown that the convergence radii are given by $\Delta_{\rm conv} = \pi T$ and $M_{\rm conv} = \sqrt{\mu^2 + (\pi T)^2}$, respectively. The difference between these two expressions can be understood physically as follows. In the BCS theory, Cooper pairing occurs close to the Fermi surface, so $\mu$ dependence enters only through the density of states, which is universal for all the GL coefficients. In the NJL model, on the other hand, the chiral condensate is a pairing between quarks and antiquarks inside the Dirac sea, so the GL coefficients depend rather nontrivially on $\mu$, as shown in Eqs.~(\ref{alpha_2})--(\ref{alpha_2n}). We also note that this difference leads to the known fact that the BCS instability appears for an infinitesimally small attractive interaction between fermions, whereas generation of the chiral condensate requires a sufficiently strong attractive interaction between quarks and antiquarks~\cite{Nambu:1961tp}. 

In this paper, we focused on the two-flavor NJL model, where only even powers of $M$ appear in the GL expansion. In the three-flavor case, on the other hand, odd powers in $M$ can also appear in the expansion \cite{Pisarski:1983ms} due to the instanton-induced interaction (or Kobayashi--Maskawa--'t Hooft interaction) \cite{Kobayashi:1970ji,tHooft:1976rip,tHooft:1976snw}. It would be interesting to see how the radius of convergence may be affected by such terms.
One can also ask about the radius of convergence of the GL theory for the chiral phase transition in QCD {\it per se}. For the mean-field approximation to be well justified, one can take the large-$N_{\rm c}$ limit~\cite{tHooft:1973alw, Witten:1979kh}.
Because the radius of convergence for the dynamical quark mass in the NJL model above does not depend on the details of the model, such as the four-fermion coupling constant $G$ and cutoff $\Lambda$,%
\footnote{Strictly speaking, this claim is valid for the Schwinger regularization, but not necessarily for the cutoff method. In the latter case, one can show that this radius of convergence holds whenever $T\geq\mu/\pi$ or $T\leq \frac1\pi \sqrt{\frac12 \Lambda^2 - \mu^2}$, and it turns out that one of these conditions is always satisfied when $T < T_{\rm c}$. This need not be the case, however, for other values of $\Lambda$ and $G$. See Appendix \ref{appendix:RoC_cut_finite_lambda}.}
one may conjecture that large-$N_{\rm c}$ QCD has the same radius of convergence, $M_{\rm conv} = \sqrt{\mu^2 + (\pi T)^2}$.%
\footnote{The chiral phase transition is either first or second order for large-$N_{\rm c}$ QCD in the chiral limit, depending on whether it coincides with the deconfinement transition~\cite{Hidaka:2011jj}. For the former case, the jump in $M$ at the phase transition has to be smaller than the radius of convergence in order for the GL solution to converge.} 

It would be interesting to extend our approach to the diquark condensate for color superconductivity~\cite{Iida:2000ha}, the interplay between chiral and diquark condensates~\cite{Hatsuda:2006ps}, the inhomogeneous condensates like Fulde--Ferrell--Larkin--Ovchinnikov (FFLO) pairing~\cite{Fulde1964, Larkin1964} or inhomogeneous chiral condensates~\cite{Buballa:2014tba}, and other orders in condensed matter systems. We defer these questions to future work.

\section*{Acknowledgment}

We thank T.~Brauner for useful correspondence and comments on the manuscript.
N.~Y.~is supported in part by the Keio Institute of Pure and Applied Sciences (KiPAS) project at Keio University and JSPS KAKENHI Grant Numbers JP19K03852 and JP22H01216. 
Furthermore, we acknowledge support from JSPS through the JSPS Summer Program 2023, where this collaboration was initiated.

\vspace{0.2cm}
\noindent

\let\doi\relax

\bibliographystyle{elsarticle-num} 
\bibliography{GL_ref}

\begin{thebibliography}{10}
\expandafter\ifx\csname url\endcsname\relax
  \def\url#1{\texttt{#1}}\fi
\expandafter\ifx\csname urlprefix\endcsname\relax\def\urlprefix{URL }\fi
\expandafter\ifx\csname href\endcsname\relax
  \def\href#1#2{#2} \def\path#1{#1}\fi

\bibitem{Dyson:1952tj}
F.~J. Dyson, {Divergence of perturbation theory in quantum electrodynamics}, Phys. Rev. 85 (1952) 631--632.
\newblock \href {https://doi.org/10.1103/PhysRev.85.631} {\path{doi:10.1103/PhysRev.85.631}}.

\bibitem{Hurst:1952zk}
C.~A. Hurst, {An example of a divergent perturbation expansion in field theory}, Proc. Cambridge Phil. Soc. 48 (1952) 625.
\newblock \href {https://doi.org/10.1017/S0305004100076416} {\path{doi:10.1017/S0305004100076416}}.

\bibitem{tHooft:1977xjm}
G.~'t~Hooft, {Can We Make Sense Out of Quantum Chromodynamics?}, Subnucl. Ser. 15 (1979) 943.

\bibitem{Aniceto:2018bis}
I.~Aniceto, G.~Basar, R.~Schiappa, {A Primer on Resurgent Transseries and Their Asymptotics}, Phys. Rept. 809 (2019) 1--135.
\newblock \href {http://arxiv.org/abs/1802.10441} {\path{arXiv:1802.10441}}, \href {https://doi.org/10.1016/j.physrep.2019.02.003} {\path{doi:10.1016/j.physrep.2019.02.003}}.

\bibitem{Grozdanov:2019kge}
S.~Grozdanov, P.~K. Kovtun, A.~O. Starinets, P.~Tadi\'c, {Convergence of the Gradient Expansion in Hydrodynamics}, Phys. Rev. Lett. 122~(25) (2019) 251601.
\newblock \href {http://arxiv.org/abs/1904.01018} {\path{arXiv:1904.01018}}, \href {https://doi.org/10.1103/PhysRevLett.122.251601} {\path{doi:10.1103/PhysRevLett.122.251601}}.

\bibitem{Cartwright:2021qpp}
C.~Cartwright, M.~G. Amano, M.~Kaminski, J.~Noronha, E.~Speranza, {Convergence of hydrodynamics in a rotating strongly coupled plasma}, Phys. Rev. D 108~(4) (2023) 046014.
\newblock \href {http://arxiv.org/abs/2112.10781} {\path{arXiv:2112.10781}}, \href {https://doi.org/10.1103/PhysRevD.108.046014} {\path{doi:10.1103/PhysRevD.108.046014}}.

\bibitem{Nambu:1961tp}
Y.~Nambu, G.~Jona-Lasinio, {Dynamical Model of Elementary Particles Based on an Analogy with Superconductivity. 1.}, Phys. Rev. 122 (1961) 345--358.
\newblock \href {https://doi.org/10.1103/PhysRev.122.345} {\path{doi:10.1103/PhysRev.122.345}}.

\bibitem{Gorkov1959}
L.~P. Gor’kov, Microscopic derivation of the ginzburg-landau equations in the theory of superconductivity, Sov. Phys. JETP 9~(6) (1959) 1364--1367.

\bibitem{ADG}
A.~Abrikosov, I.~Dzyaloshinskii, L.~Gor’kov, Methods of Quantum Field Theory in Statistical Physics, Dover, New York, 1975.

\bibitem{Brauner}
T.~Brauner, Asymptotic expansion of singular integrals (unpublished notes), \url{https://sites.google.com/site/braunercz/notes}.

\bibitem{Laine2016}
M.~Laine, A.~Vuorinen, Basics of thermal field theory, Lect. Notes Phys 925~(1) (2016) 1701--01554.

\bibitem{Klevansky:1992qe}
S.~P. Klevansky, {The Nambu-Jona-Lasinio model of quantum chromodynamics}, Rev. Mod. Phys. 64 (1992) 649--708.
\newblock \href {https://doi.org/10.1103/RevModPhys.64.649} {\path{doi:10.1103/RevModPhys.64.649}}.

\bibitem{Hatsuda:1994pi}
T.~Hatsuda, T.~Kunihiro, {QCD phenomenology based on a chiral effective Lagrangian}, Phys. Rept. 247 (1994) 221--367.
\newblock \href {http://arxiv.org/abs/hep-ph/9401310} {\path{arXiv:hep-ph/9401310}}, \href {https://doi.org/10.1016/0370-1573(94)90022-1} {\path{doi:10.1016/0370-1573(94)90022-1}}.

\bibitem{Buballa:2003qv}
M.~Buballa, {NJL model analysis of quark matter at large density}, Phys. Rept. 407 (2005) 205--376.
\newblock \href {http://arxiv.org/abs/hep-ph/0402234} {\path{arXiv:hep-ph/0402234}}, \href {https://doi.org/10.1016/j.physrep.2004.11.004} {\path{doi:10.1016/j.physrep.2004.11.004}}.

\bibitem{Buballa:2014tba}
M.~Buballa, S.~Carignano, {Inhomogeneous chiral condensates}, Prog. Part. Nucl. Phys. 81 (2015) 39--96.
\newblock \href {http://arxiv.org/abs/1406.1367} {\path{arXiv:1406.1367}}, \href {https://doi.org/10.1016/j.ppnp.2014.11.001} {\path{doi:10.1016/j.ppnp.2014.11.001}}.

\bibitem{Schwinger:1951nm}
J.~S. Schwinger, {On gauge invariance and vacuum polarization}, Phys. Rev. 82 (1951) 664--679.
\newblock \href {https://doi.org/10.1103/PhysRev.82.664} {\path{doi:10.1103/PhysRev.82.664}}.

\bibitem{Yamamoto:2007ah}
N.~Yamamoto, M.~Tachibana, T.~Hatsuda, G.~Baym, {Phase structure, collective modes, and the axial anomaly in dense QCD}, Phys. Rev. D 76 (2007) 074001.
\newblock \href {http://arxiv.org/abs/0704.2654} {\path{arXiv:0704.2654}}, \href {https://doi.org/10.1103/PhysRevD.76.074001} {\path{doi:10.1103/PhysRevD.76.074001}}.

\bibitem{Apostol1976}
T.~M. Apostol, Introduction to Analytic Number Theory, Springer, New York, 1976.
\newblock \href {https://doi.org/10.1007/978-1-4757-5579-4_13} {\path{doi:10.1007/978-1-4757-5579-4_13}}.

\bibitem{Gyory}
W.~Gyory, Phase transitions and thermal stability of the magnetic dual chiral density wave phase in cold, dense {QCD}, Ph.D. thesis, CUNY Graduate Center (2023).

\bibitem{Nickel:2009wj}
D.~Nickel, {Inhomogeneous phases in the Nambu-Jona-Lasino and quark-meson model}, Phys. Rev. D 80 (2009) 074025.
\newblock \href {http://arxiv.org/abs/0906.5295} {\path{arXiv:0906.5295}}, \href {https://doi.org/10.1103/PhysRevD.80.074025} {\path{doi:10.1103/PhysRevD.80.074025}}.

\bibitem{Pisarski:1983ms}
R.~D. Pisarski, F.~Wilczek, {Remarks on the Chiral Phase Transition in Chromodynamics}, Phys. Rev. D 29 (1984) 338--341.
\newblock \href {https://doi.org/10.1103/PhysRevD.29.338} {\path{doi:10.1103/PhysRevD.29.338}}.

\bibitem{Kobayashi:1970ji}
M.~Kobayashi, T.~Maskawa, {Chiral symmetry and eta-x mixing}, Prog. Theor. Phys. 44 (1970) 1422--1424.
\newblock \href {https://doi.org/10.1143/PTP.44.1422} {\path{doi:10.1143/PTP.44.1422}}.

\bibitem{tHooft:1976rip}
G.~'t~Hooft, {Symmetry Breaking Through Bell-Jackiw Anomalies}, Phys. Rev. Lett. 37 (1976) 8--11.
\newblock \href {https://doi.org/10.1103/PhysRevLett.37.8} {\path{doi:10.1103/PhysRevLett.37.8}}.

\bibitem{tHooft:1976snw}
G.~'t~Hooft, {Computation of the Quantum Effects Due to a Four-Dimensional Pseudoparticle}, Phys. Rev. D 14 (1976) 3432--3450, [Erratum: Phys.Rev.D 18, 2199 (1978)].
\newblock \href {https://doi.org/10.1103/PhysRevD.14.3432} {\path{doi:10.1103/PhysRevD.14.3432}}.

\bibitem{tHooft:1973alw}
G.~'t~Hooft, {A Planar Diagram Theory for Strong Interactions}, Nucl. Phys. B 72 (1974) 461.
\newblock \href {https://doi.org/10.1016/0550-3213(74)90154-0} {\path{doi:10.1016/0550-3213(74)90154-0}}.

\bibitem{Witten:1979kh}
E.~Witten, {Baryons in the 1/n Expansion}, Nucl. Phys. B 160 (1979) 57--115.
\newblock \href {https://doi.org/10.1016/0550-3213(79)90232-3} {\path{doi:10.1016/0550-3213(79)90232-3}}.

\bibitem{Hidaka:2011jj}
Y.~Hidaka, N.~Yamamoto, {No-Go Theorem for Critical Phenomena in Large-Nc QCD}, Phys. Rev. Lett. 108 (2012) 121601.
\newblock \href {http://arxiv.org/abs/1110.3044} {\path{arXiv:1110.3044}}, \href {https://doi.org/10.1103/PhysRevLett.108.121601} {\path{doi:10.1103/PhysRevLett.108.121601}}.

\bibitem{Iida:2000ha}
K.~Iida, G.~Baym, {The Superfluid phases of quark matter: Ginzburg-Landau theory and color neutrality}, Phys. Rev. D 63 (2001) 074018, [Erratum: Phys.Rev.D 66, 059903 (2002)].
\newblock \href {http://arxiv.org/abs/hep-ph/0011229} {\path{arXiv:hep-ph/0011229}}, \href {https://doi.org/10.1103/PhysRevD.63.074018} {\path{doi:10.1103/PhysRevD.63.074018}}.

\bibitem{Hatsuda:2006ps}
T.~Hatsuda, M.~Tachibana, N.~Yamamoto, G.~Baym, {New critical point induced by the axial anomaly in dense QCD}, Phys. Rev. Lett. 97 (2006) 122001.
\newblock \href {http://arxiv.org/abs/hep-ph/0605018} {\path{arXiv:hep-ph/0605018}}, \href {https://doi.org/10.1103/PhysRevLett.97.122001} {\path{doi:10.1103/PhysRevLett.97.122001}}.

\bibitem{Fulde1964}
P.~Fulde, R.~A. Ferrell, Superconductivity in a strong spin-exchange field, Physical Review 135~(3A) (1964) A550.

\bibitem{Larkin1964}
A.~Larkin, Y.~N. Ovchinnikov, Nonuniform state of superconductors, Zh. Eksperim. i Teor. Fiz. 47 (1964) 1136.

\bibitem{Nakano:2004cd}
E.~Nakano, T.~Tatsumi, {Chiral symmetry and density wave in quark matter}, Phys. Rev. D 71 (2005) 114006.
\newblock \href {http://arxiv.org/abs/hep-ph/0411350} {\path{arXiv:hep-ph/0411350}}, \href {https://doi.org/10.1103/PhysRevD.71.114006} {\path{doi:10.1103/PhysRevD.71.114006}}.

\bibitem{Gyory:2022hnv}
W.~Gyory, V.~de~la Incera, {Phase transitions and resilience of the magnetic dual chiral density wave phase at finite temperature and density}, Phys. Rev. D 106~(1) (2022) 016011.
\newblock \href {http://arxiv.org/abs/2203.14209} {\path{arXiv:2203.14209}}, \href {https://doi.org/10.1103/PhysRevD.106.016011} {\path{doi:10.1103/PhysRevD.106.016011}}.

\end{thebibliography}

\appendix

\section{GL coefficients in the NJL model from the proper time method}
\label{appendix:cc_proper_time}

The general idea of the proper time regularization amounts to making the following replacement by introducing a UV cutoff $\Lambda$:
\begin{equation}
    |E_{\bm k}|^{-2n} = \frac{1}{\Gamma(n)}\int_{0}^\infty
    {d s}\, s^{n-1} e^{-s E_{\bm k}^2} \to 
    \frac{1}{\Gamma(n)} \int_{1 / \Lambda^2}^\infty
    {d s}\, s^{n-1} e^{-s E_{\bm k}^2}\,.
\end{equation}
For our purpose, we need the $n=-\frac{1}{2}$ case, 
\begin{equation}
\label{reg}
    |E_{\bm k}| \to - \frac1{2\sqrt\pi}
    \int_{1 / \Lambda^2}^\infty
    \frac{d s}{s^{3/2}} e^{-s E_{\bm k}^2}\,,
\end{equation}
for only the first term of Eq.~(\ref{omega_CC_pre_reg}), where we used $\Gamma(-\frac{1}{2})=-2\sqrt{\pi}$. 
The replacement in Eq.~(\ref{reg}) arises more naturally when deriving the free energy $\Omega$ in the proper time framework, where the divergences appear as integrals of the form $\int_0^\infty \frac{d s} {s^{3/2}} e^{-sE_{\bm k}^2}$ (see, e.g.\ Appendix A of Ref.~\cite{Nakano:2004cd}). 

We can now express the free energy as a well-defined quantity,
\begin{align}
    \Omega 
    &= \frac{M^2}{4G} - \frac {N_{\rm f} N_{\rm c}} {\pi^2}
    T^4 J_2^\text{Schw}
    (\beta M, \beta \mu; \beta \Lambda)\,,
    \label{omega_CC_J2}
\end{align}
where
\begin{equation}
    J_{\ell}^\text{Schw}(x, y; L)
    = \int_0^\infty d t \, t^{\ell} 
    \left[
        \left(- \frac1{2\sqrt\pi}
        \int_{1 / L^2}^\infty \frac {d s} {s^{3/2}} e^{-s(x^2 + t^2)}
        \right)
        + \sum_{\zeta=\pm} 
        \ln
        \left(
            1+e^{-(\sqrt{x^2 + t^2} + \zeta y)}
        \right)
    \right].
\end{equation}

Let us introduce the function
\begin{equation}
    F(z) = 
    -\frac L {\sqrt \pi}
    e^{- (z / L)^2}
    + z \operatorname{erfc}(z / L)
    + \sum_{\zeta = \pm 1} \ln
    \left(
        1 + e^{- z + \zeta y}
    \right),
\end{equation}
where $\operatorname{erfc}(x) = \frac{2}{\sqrt{\pi}}\int_x^{\infty}{d}t\, {e}^{-t^2}$ is the complementary error function, and we have suppressed the dependence on $L$ and $y$ to reduce clutter. 
One can show that $t^2 F(\sqrt{x^2 + t^2})$ is precisely the integrand of $J_2^\text{Schw}(x, y; L)$, and this allows us to write $J_2^\text{Schw} = \frac12 \int_{- \infty}^{+ \infty} d t \, \allowbreak t^2 F(\sqrt{x^2 + t^2})$. Note also that $F(z)$ is an even function of $z$ that vanishes rapidly at $\pm \infty$, which are important properties in what follows.

We then need to compute $\partial_{x^2}^n F(\sqrt{x^2 + t^2}) \big|_{x = 0} = 2^{-n}(t^{-1} \partial_t)^n F(t)$. The $n = 1$ case combined with an integration by parts immediately yields
\begin{equation}
    c_2 = -\frac 1 4 \int_{- \infty}^{+ \infty} d t\, F(t)\,.
\end{equation}
For the higher coefficients, we can put $(t^{-1} \partial_t)^n$ into the form $t^{-1} \partial_t^{2n + 1}$ using repeated integration by parts. One can show that \cite{Gyory:2022hnv, Gyory}
\begin{equation}
    \label{x_to_t_derivatives_J2}
    \partial_{x^2}^n
    \int_{- \infty}^{+ \infty} d t \, t^2
    F\big(\sqrt{x^2 + t^2}\big) \Big|_{x = 0}
    =
    - \frac{1} {2^{n}(2n - 4)!!}
    \int_{- \infty}^{+ \infty} d t \, t^{-1}
    \partial_t^{2n - 3} F(t)
    \qquad \quad n \geq 2.
\end{equation}
The above integrals $\int_{- \infty}^{+ \infty} d t F(t)$ and $\int_{-\infty}^{+ \infty} d t\, t^{-1} \partial_t^n F(t)$ can be analytically performed for all $n$~\cite{Gyory}:
\newcommand{\cl}{.55\textwidth}
\begin{align}
    &
    \int_{- \infty}^{+ \infty} d t F(t)
    =
    - \frac12 L^2 + y^2 + \frac13 \pi^2\,,
    \\
    &
    \int_{- \infty}^{+ \infty} d t \, t^{-1}
        \partial_t^n F(t)
        \nonumber \\
    &
    =\begin{dcases}
        \mathmakebox[\cl][l]{
            2 \ln
            \left( \frac L {4 \pi e^{\gamma / 2}} \right)
            - 2 \operatorname{Re} \psi
            \left(
            \frac12 + i\frac y {2\pi}
            \right)
            }
            \qquad \qquad n = 1 \\
        \mathmakebox[\cl][l]{
            \frac{(-2)^{(n + 1) / 2} (n - 3)!!} {L^{n - 1}}
            + \frac {2 (-1)^{(n + 1) / 2}} 
            {(2 \pi)^{n - 1}}
            \operatorname{Re} 
            \psi^{(n - 1)}\left(
            \frac12 + i\frac y {2 \pi}
            \right)
            } 
            \qquad \qquad n = 3,5,\dots\,.
    \end{dcases}
    \label{integral_formula_2}
\end{align}
We therefore have
\begin{align}
    c_2
    &= \frac14 
    \left( 
        \frac12 L^2 - y^2 - \frac13 \pi^2
    \right)\,,
    \label{a_2_J2_Schw}
    \\
    c_{4}
    &= \frac18
    \left[
        - \ln 
        \left(
            \frac L
            {4 \pi e^{\gamma / 2}}
        \right)
        + \operatorname{Re}
        \psi
        \left(
            \frac12 + i\frac y {2 \pi}
        \right)
    \right]\,,
\end{align}
and
\begin{align}
    c_{2n \geq 6}
    &= c_{2n}^\text{vac} + c_{2n}^\text{med},
    \label{a_2n_J2_Schw}
    \\
    c_{2n}^\text{vac}
    &= \frac {(-1)^n} {n! \, (2n - 4)!!} 
    \bigg[ 
    \frac{(2n - 6)!!} {4 L^{2n - 4}}
    \bigg]\,,
    \label{a_2n_vac}
    \\
    c_{2n}^\text{med}
    &= \frac {(-1)^n} {n! \, (2n - 4)!!} 
    \bigg[ 
    \frac {1} {2^n (2 \pi)^{2n - 4}}
    \operatorname{Re} 
    \psi^{(2n - 4)}
    \left(
        \frac12 + i\frac y {2 \pi}
    \right) 
    \bigg]\,.
    \label{a_2n_med}
\end{align}

It now follows from Eq.~(\ref{omega_CC_J2}) that the GL coefficients are given by
\begin{equation}
    \alpha_{2n}
    = \frac {\delta_{2,n}} {4G} 
    - \frac {N_{\rm f} N_{\rm c}} {\pi^2}
    T^{4 - 2n} c_{2n}
    \Big|_{y \to \beta \mu, L \to \beta \Lambda}\,.
\end{equation}
Explicitly, 
\begin{align}
    \alpha_2
    &= \frac 1 {4G} - \frac {N_{\rm f} N_{\rm c}} {\pi^2}
    \frac 1 4
    \left(
        \frac {\Lambda^2} 2 
        - \mu^2
        - \frac 1 3 \pi^2 T^2
    \right)\,,
    \\
   \label{alpha_4} 
    \alpha_4
    &= - \frac{N_{\rm f} N_{\rm c}} {\pi^2} \frac 1 8
    \left[
        - \ln\left(
            \frac {\Lambda} 
            {4 \pi e^{\gamma / 2} T}
        \right)
        + \operatorname{Re} \psi \left(
            \frac 1 2 + i
            \frac {\mu} {2 \pi T}
        \right)
    \right]\,,
    \\
    \alpha_{2n \geq 6}
    &= - \frac{N_{\rm f} N_{\rm c}} {\pi^2}
    \frac {(-1)^n} {n! \, (2n - 4)!!}
    \bigg[
        \frac {(2n - 6)!!} 
        {4 \Lambda^{2n - 4}} 
    + \frac 1 {2^n (2 \pi T)^{2n - 4}}
    \operatorname{Re} 
    \psi^{(2n - 4)}
    \left(
    \frac12 + i\frac {\mu} {2 \pi T}
    \right) 
    \bigg]\,.
\end{align}
Note that the only differences between these coefficients and those given in Eqs.~(\ref{alpha_2})--(\ref{alpha_2n}) are in the terms involving $\Lambda$, as expected. We also note that these coefficients can be regarded as a special case of those derived in Refs.~\cite{Gyory, Gyory:2022hnv} for the case of an inhomogeneous chiral condensate in a magnetic field, after setting $B = 0$ and considering only terms in the GL expansion that do not contain any gradients.

Let us now focus on the coefficients $c_{2n \geq 6}$ and consider the convergence of the corresponding series. Since each coefficient is a sum of two parts, $c_{2n \geq 6} = c_{2n}^\text{vac} + c_{2n}^\text{med}$, the series for $J_2^\text{Schw}(x, y; L)$ can be separated into two subseries,
\begin{align}
    J_2^\text{Schw}(x, y; L)
    &= c_2 x^2 + c_4 x^4 
    + c_6 x^6 + \cdots
    \label{J2_Schw_decomp} \nonumber \\
    &= c_2 x^2 + c_4 x^4
    \nonumber \\
    &\quad + (c_6^\text{vac} x^6 + c_8^\text{vac} x^8 + \cdots) 
    \nonumber \\
    &\quad + (c_6^\text{med} x^6 + c_8^\text{med} x^8 + \cdots).
\end{align}
This rearrangement of terms is justified because rearrangements can only affect a conditionally convergent series (or more accurately, a series for which some rearrangement converges conditionally), and power series can only converge conditionally at their exact radius of convergence. Thus, rearranging a power series cannot change its convergence radius. It is easy to check that the first subseries, whose coefficients are $c_{2n}^\text{vac}$, has infinite radius of convergence, e.g.\ by applying the ratio test. Therefore, the convergence radius of the original series is determined by that of the second subseries, whose coefficients are $c_{2n}^\text{med}$. But $c_{2n\geq 6}^\text{med}$ are precisely the coefficients $c_{2n\geq 6}$ that were calculated in Sect.~\ref{subsec:CC_coef_formulas}  using the momentum cutoff, so the radius of convergence here is the same as in the previous case.

\section{Convergence radius for the GL expansion in the NJL model}

When computing the GL coefficients for the NJL model in Sect.~\ref{subsec:CC_coef_formulas}, we approximated $\Lambda$ as being very large compared to other characteristic quantities of the system, resulting in the coefficients given by Eqs.~(\ref{alpha_2})--(\ref{alpha_2n}). Let us denote the radius of convergence associated with these coefficients by $M_\text{conv}^{\text{cut,}\Lambda \to \infty}$. One can also consider the radius of convergence of the GL expansion obtained using cutoff regularization, but without taking the limit $\Lambda \rightarrow \infty$ (although we have not found simple analytical formulas for the coefficients in this case). Let us denote the latter radius of convergence by $M_\text{conv}^{\text{cut}}$. Finally, let us denote by $M_\text{conv}^\text{Schw}$ the radius of convergence of the GL expansion when using the Schwinger proper time regularization scheme. 

Using the above notation, we can summarize the previous results as follows:
\begin{equation}
    M_\text{conv}^\text{Schw} 
    \overset{({\rm i})}{=}
    M_\text{conv}^{\text{cut,}\Lambda \to \infty} 
    \overset{({\rm ii})}{\geq}
    \sqrt{\mu^2 + (\pi T)^2}.
\end{equation}
(i) was shown at the end of Appendix \ref{appendix:cc_proper_time}, and (ii) was shown at the end of Sect.~\ref{subsec:RoC_CC}. In the first part of this appendix, we show that equality holds in (ii), i.e.\ $M_\text{conv}^{\text{cut,}\Lambda \to \infty} = \sqrt{\mu^2 + (\pi T)^2}$. In the second part of this appendix, we consider the case of cutoff regularization without assuming $\Lambda \to \infty$, and we show that $M_\text{conv}^{\text{cut}} \geq \sqrt{\mu^2 + (\pi T)^2}$ over the region in the $\mu$-$T$ plane where $M > 0$, after fixing $\Lambda$ and $G$ such that $M_0 = 300\,\text{MeV}$ and $f_\pi = 88\,\text{MeV}$.

\subsection{Cutoff regularization with \texorpdfstring{$\Lambda \to \infty$}{lambda to infinity}}
\label{appendix:Cauchy-Hadamard}

According to the Cauchy--Hadamard theorem, the radius of convergence of the series $c_2 x^2 + c_4 x^4 + \cdots$ is given by
\begin{equation}
\label{Cauchy-Hadamard}
    (x_\text{conv})^{-1}
    = \limsup_{n \to \infty}
    \sqrt[2n]{|c_{2n}|}\,.
\end{equation}
Applying this theorem to $c_{2n}$ given by Eq.~(\ref{c2n}), and using the formula $\psi^{(n)}(x) = (-1)^{n + 1} n! \, \zeta(n + 1, x)$ and the fact that $\sqrt[n]{|a|} \to 1$ as $n \to \infty$ for any $a \neq 0$, we find
\begin{equation}
    (x_\text{conv})^{-1}
    = \frac 1 {2 \pi} 
    \limsup_{n \to \infty}
    \sqrt[2n]
    {\frac {(2n - 5)!!} {(2n)!!}}
    \sqrt[2n]
    {\Bigg| \operatorname{Re} 
    \sum_{j = 0}^\infty
    \frac 1 
    {(j + \tfrac12 + i\tfrac y {2 \pi})^{2n - 3}}
    \Bigg|}\,.
\end{equation}
It is easy to show that the first factor in the limit approaches $1$.
For the second factor, the terms except for $j=0$ in the sum over $j$ become negligible as $n \to \infty$, so we have
\begin{align}
    (x_\text{conv})^{-1}
    &= \frac {|z|}{2 \pi} 
    \limsup_{n \to \infty}
    \sqrt[2n]
    {\big| \operatorname{Re} 
    (\hat z^{2n - 3})
    \big|}\,, 
\end{align}
where $z = 1/\left({\tfrac12 + i\tfrac y {2 \pi}} \right)$ and $\hat z = z / |z| = e^{2\pi i\phi}$ is a complex phase of unit magnitude. The result will follow if we can show that the remaining lim sup evaluates to unity. First, we have $\sqrt[2n]{|\operatorname{Re}(\hat z^{2n - 3})|} \leq \sqrt[2n]{|(\hat z^{2n - 3})|} = 1$, so we must only show that $\sqrt[2n]{|\operatorname{Re}(\hat z^{2n - 3})|}$ contains a subsequence converging to $1$. If $\phi$ is rational, then $\hat z^{2n - 3}$ will cycle through finitely many values infinitely many times. Letting $a = |\operatorname{Re}(\hat z)|$, there is a subsequence $\sqrt[2n]{a}$, which converges to $1$. If $\phi$ is irrational, then the set of points $\{e^{2\pi i\phi n} \, | \, n \in \mathbb N \}$ is dense in the unit circle, and hence so is the set $\{ e^{2\pi i\phi (2n - 3)} \, | \, n \in \mathbb N \}$. Thus we can choose a subsequence $\{a_n\}$ of $e^{2\pi i\phi (2n - 3)}$ whose elements all have real part at least $\frac{1}{2}$, and then $\sqrt[2n]{|\operatorname{Re}(a_n)|} \to 1$.

\subsection{Cutoff regularization with finite \texorpdfstring{$\Lambda$}{lambda}}
\label{appendix:RoC_cut_finite_lambda}

Starting from Eq.~(\ref{d_J2_cut}) and repeating the calculation without taking $\lambda\to\infty$, we find
\begin{equation}
    \label{c_2n_finite_lambda}
    c_{2n\geq 6}
    = \frac{2 (-1)^{n - 1}} n 
    \operatorname{Re} \sum_{k = 0}^\infty
    \frac 1 {A_k^{2n - 3}} 
    \int_0^{\lambda / A_k}
    {d}s \frac{s^2}{(1 + s^2)^n},
\end{equation}
where we have defined $A_k := (2k + 1)\pi + {i} y$.
Because $\lambda / A_k$ is complex, the integral in Eq.~(\ref{c_2n_finite_lambda}) must now be interpreted as a contour integral. Let $C_k$ be some contour that starts at the origin, ends at $\lambda / A_k$, and avoids the poles at $s = \pm i$; we will consider two different choices of $C_k$. Let $|C_k|$ denote the length of the contour $C_k$.

Applying the Cauchy--Hadamard theorem~(\ref{Cauchy-Hadamard}), we have
\begin{align}
    (x_\text{conv})^{-1}
    &\leq \limsup_{n \to \infty}
    \sqrt[2n]{\sum_{k = 0}^\infty
    \frac 1 {|A_k|^{2n - 3}} 
    |C_k| 
    \max_{s \in C_k} 
    \left| \frac{s^2}{(1 + s^2)^n} \right|}\,.
    \label{lim_sup_Ck}
\end{align}
Recall that Eq.~(\ref{lim_sup_Ck}) holds for any valid choice of contours $C_k$. Let us first consider contours that follow the straight line from the origin to $\lambda / A_k$. If $y \leq \pi$, then $y \leq (2k + 1)\pi$ for all $k$, and it is easy to show that $|1 + s^2| \geq 1$ for all $s \in C_k$. We therefore have
\begin{align}
    (x_\text{conv})^{-1}
    &\leq \limsup_{n \to \infty}
    \sqrt[2n]{\sum_{k = 0}^\infty
    \frac 1 {|A_k|^{2n - 3}} 
    \left|\frac\lambda{A_k}\right| 
    \left|\frac\lambda{A_k}\right|^2} \nonumber \\
    &= \frac1{|A_0|}\,,
\end{align}
which shows that $x_\text{conv} \geq \sqrt{\pi^2 + y^2}$ if $y \leq \pi$. 

More generally, for any $y \in \mathbb R$, one can derive the weaker lower bound
\begin{equation}
    \min_{y \in C_k} 
    |1 + y^2| 
    \geq 2\frac{(2k + 1)\pi y}{|A_k|^2}\,,
\end{equation}
from which we find
\begin{align}
    (x_\text{conv})^{-1}
    &\leq \limsup_{n \to \infty}
    \sqrt[2n]{\sum_{k = 0}^\infty
    \frac 1 {|A_k|^{2n - 3}} 
    \left|\frac\lambda{A_k}\right| 
    \left|\frac\lambda{A_k}\right|^2
    \left(
        \frac{|A_k|^2}{2(2k + 1)\pi y}
    \right)^n}
    \nonumber \\
    &= \frac1{\sqrt{2\pi y}}\,.
\end{align}
Thus, we always have $x_\text{conv} \geq \sqrt{2\pi y}$, or equivalently, $M_\text{conv} \geq \sqrt{2\pi\mu T}$. It is easy to check that the previous lower bound $x_\text{conv} \geq \sqrt{\pi^2 + y^2}$, which holds (at least) if $y \leq \pi$, is always an improvement over $\sqrt{2\pi y}$ (except precisely at $y = \pi$, where the two bounds are equal).

Finally, by considering a different choice of the contour $C_0$, one can show that the improved lower bound $x_\text{conv} \geq \sqrt{\pi^2 + y^2}$ holds over a larger region of parameter space. Let $C_0$ now be the contour that runs first along the positive real axis from $0$ to $|\lambda / A_0|$, and then along a circular arc of constant radius from $|\lambda / A_0|$ to $\lambda / A_0$ (we let $C_k$ for $k > 0$ be the same as before). It is then easy to show that if $|\lambda / A_0| \geq \sqrt2$, then again $|1 + s^2| \geq 1$ for all $s \in C_0$. We also have $|C_k| \leq (1 + \pi/2)|\lambda / A_k|$ for all $k$, from which we find
\begin{align}
    (x_\text{conv})^{-1}
    &\leq \limsup_{n \to \infty}
    \sqrt[2n]{\sum_{k = 0}^\infty
    \frac 1 {|A_k|^{2n - 3}}
    \left( 1 + \frac{\pi}{2} \right)
    \left|\frac\lambda{A_k}\right| 
    \left|\frac\lambda{A_k}\right|^2
    \max_{s \in C_k}\frac1{|1 + s^2|^n}}
    \nonumber \\
    &= \frac1{|A_0|}\,.
\end{align}

We have shown that $x_\text{conv} \geq \sqrt{2\pi y}$ for all $y \in \mathbb R$, and that the improved lower bound $x_\text{conv} \geq \sqrt{\pi^2 + y^2}$ holds if $y \leq \pi$ or $|\lambda / A_0| \geq \sqrt2$ (in fact, these conditions are sufficient, but not necessary). The condition $|\lambda / A_0| \geq \sqrt2$ is equivalent to $y \leq \sqrt{\frac12 \lambda^2 - \pi^2}$, and it follows that
\begin{equation}
\label{M_conv_cut_finite_lambda}
\left.
\begin{aligned}
T &\geq \mu/\pi \\
& \text{or} \\
T &\leq \tfrac1\pi \sqrt{\tfrac12 \Lambda^2 - \mu^2}   \\
\end{aligned}
\quad \right\} \implies
M_\text{conv} \geq \sqrt{\mu^2 + (\pi T)^2}.
\end{equation}

\begin{figure}
    \centering
    \includegraphics[width = .5 \textwidth]{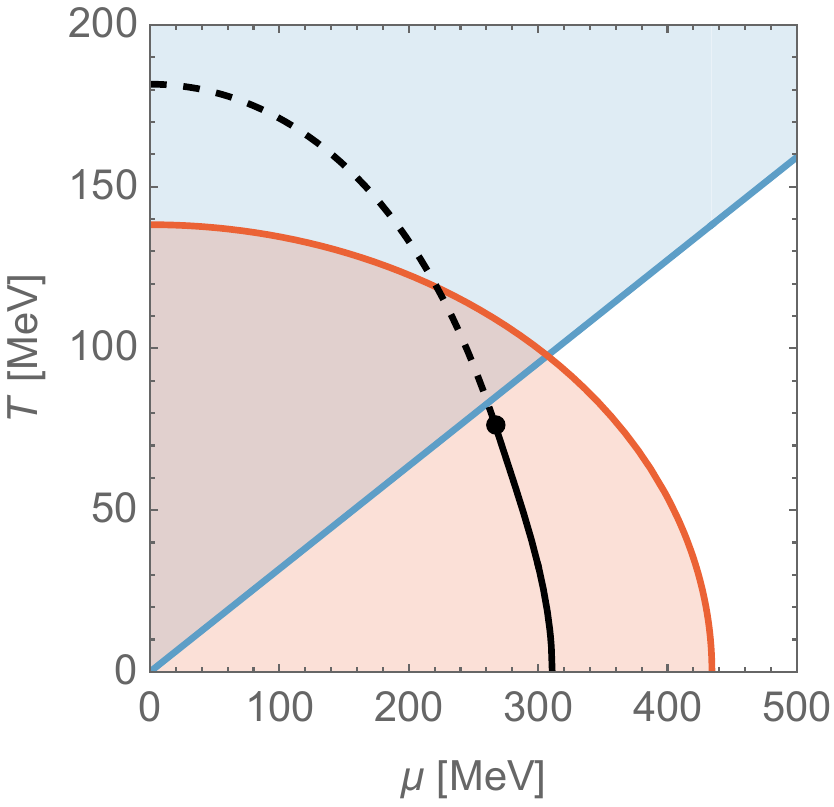}
    \caption{Convergence behavior of the GL expansion using cutoff regularization without assuming $\Lambda = \infty$. In the shaded regions corresponding to the conditions given in Eq.~(\ref{M_conv_cut_finite_lambda}), we have $M_\text{conv}^\text{cut} \geq \sqrt{\mu^2 + (\pi T)^2}$, and in the unshaded region we have the weaker lower bound $M_\text{conv}^\text{cut} \geq \sqrt{2\pi\mu T}$.  The blue shaded region corresponds to $T \geq \mu / \pi$, and the red shaded region corresponds to $T \leq \frac 1 \pi \sqrt{\frac12 \Lambda^2 - \mu^2}$. The black curve shows $T_{\rm c}$, computed by solving the gap equation numerically with cutoff regularization, with solid (dashed) lines indicating the first-order (second-order) transitions. We use $\Lambda = 614\,\text{MeV}$ and $G\Lambda^2 = 2.15$ as described in Sect.~\ref{subsec:RoC_CC}.
    }
    \label{fig:GL_Conv_CC_Finite}
\end{figure}

Fig.~\ref{fig:GL_Conv_CC_Finite} shows the region in the $\mu$-$T$ plane where the two conditions in Eq.~(\ref{M_conv_cut_finite_lambda}) hold, along with $T_{\rm c}$ computed using cutoff regularization. 

\end{document}